\documentclass[pdflatex,sn-basic]{sn-jnl}

\usepackage{graphicx}%
\usepackage{multirow}%
\usepackage{amsmath,amssymb,amsfonts}%
\usepackage{amsthm}%
\usepackage{mathrsfs}%
\usepackage[title]{appendix}%
\usepackage{xcolor}%
\usepackage{textcomp}%
\usepackage{manyfoot}%
\usepackage{booktabs}%
\usepackage{algorithm}%
\usepackage{algorithmicx}%
\usepackage{algpseudocode}%
\usepackage{listings}%

\usepackage{bm} 
\usepackage{setspace}
\usepackage{array}
\usepackage{adjustbox}
\usepackage{makecell}
\usepackage{tabularx}
\usepackage{rotating}
\usepackage{fancyhdr}
\usepackage{enumitem}

\theoremstyle{thmstyleone}%
\theoremstyle{thmstyletwo}%

\theoremstyle{thmstylethree}%

\newtheorem{problem}{Problem}

\usepackage{geometry}
\usepackage{setspace}

\begin{document}

\title[POD-based ROM for time-dependent coupled diffusion-deformation of hydrogels]{Parameter identification and uncertainty propagation of hydrogel coupled diffusion-deformation using POD-based reduced-order modeling}


\author*[1]{\fnm{Gopal} \sur{Agarwal}}\email{agarwal.gopal@columbia.edu}
\equalcont{These authors contributed equally to this work.}

\author*[2]{\fnm{Jorge-Humberto} \sur{Urrea-Quintero}}\email{jorge.urrea-quintero@tu-braunschweig.de}
\equalcont{These authors contributed equally to this work.}

\author*[2]{\fnm{Henning} \sur{Wessels}}\email{h.wessels@tu-braunschweig.de}

\author*[1]{\fnm{Thomas} \sur{Wick}}\email{wick@ifam.uni-hannover.de}

\affil[1]{\orgname{Leibniz University Hannover}, \orgdiv{Institut für Angewandte Mathematik}, \orgaddress{\street{Welfengarten 1}, \city{Hannover}, \postcode{30167}, \state{Lower Saxony}, \country{Germany}}}

\affil[2]{\orgname{Technische Universtität Braunschweig}, \orgdiv{Institute for Computational Modeling in Civil Engineering}, \orgaddress{\street{Pockelsstr. 3}, \city{Braunschweig}, \postcode{38106}, \state{Lower Saxony}, \country{Germany}}}

\affil[3]{\orgname{Leibniz University Hannover}, \orgdiv{Cluster of Excellence PhoenixD (Photonics, Optics, and Engineering - Innovation Across Disciplines)}, \orgaddress{\street{Welfengarten 1}, \city{Hannover}, \postcode{30167}, \state{Lower Saxony}, \country{Germany}}}


\abstract{This study explores reduced-order modeling for analyzing the time-dependent diffusion-deformation of hydrogels. The full-order model describing hydrogel transient behavior consists of a coupled system of partial differential equations in which chemical potential and displacements are coupled. This system is formulated in a monolithic fashion and solved using the finite element method. We employ proper orthogonal decomposition as a model order reduction approach. The reduced-order model performance is tested through a benchmark problem on hydrogel swelling and a case study simulating co-axial printing. Then, we embed the reduced-order model into an optimization loop to efficiently identify the coupled problem's material parameters using full-field data. Finally, a study is conducted on the uncertainty propagation of the material parameter.}




\maketitle

\newpage

\section{Introduction}

Hydrogels, known for their biocompatibility, high water content, and adaptable mechanical properties, are at the forefront of innovation in the fabrication of functional materials for biomedical applications \citep{Zhang2017HydrogelsScience}.
The fabrication of hydrogels has been supported by the integration of quantitative experimental techniques with computational models describing the material and mechanical properties of the fabricated hydrogel. 
A comprehensive review of available experimental methods to investigate hydrogel properties, along with an overview of some common mathematical modeling approaches describing hydrogels' mechanics and transport properties, can be found in \cite{Caccavo2018hydrogels}.
A recent review of the theories capturing hydrogel's diffusion-deformation coupled mechanisms 
can be found in \cite{Urrea2023DiffDef_hydrogels}.

Diffusion-deformation models for hydrogels combine continuum mechanics and thermodynamics with solvent diffusion physics through the balance of mass and Fick’s law of diffusion \citep{Chester2010DiffDeform,Bouklas2015Nonlinear,Liu2016TransGels}, but lack an analytical solution in general \citep{Chester2015Abaqus,Urrea2023DiffDef_hydrogels}. Hence, the model’s solution requires numerical methods, such as the Finite Element Method (FEM) \citep{Anand2020CMM_solids_Book,Urrea2023DiffDef_hydrogels}. 
Running such a full-order model (FOM) finite element simulation can become costly \citep{Bonatti2021SimBioprinting, Chirianni2024BioprintingStress}. This has hindered the widespread adoption of multiphysics modeling in hydrogel manufacturing, where many-query tasks are often performed, and several hundreds or thousands of similar runs are often necessary. Typical many-query tasks are material parameter identification \citep{Zou2018PGD_Ident_Biomechs}, uncertainty quantification \citep{Pagani2021EnablingUQ_POD}, process optimization, model-based 3D printing monitoring, or feedback printing error compensation technologies.\citep{Lassila2014MOR,Strazzullo2020ROMControl}.

Model-order reduction (MOR) offers a promising solution to the challenges posed by the computational burden of multiphysics models. 
By using a ROM, we can save on computation resources while still capturing the dominant behaviors of the system. 
ROMs can be tailored to specific regions of interest, ensuring accuracy where it is most crucial. 
ROMs can bridge the gap between the detailed insights provided by multiphysics models and the practical necessities of the many-query tasks mentioned above; see, for example, \cite{Chen2015ROMUQ}, \cite{Strazzullo2020ROMControl}, and \cite{Pagani2021EnablingUQ_POD}. 

In the MOR literature, Proper Orthogonal Decomposition (POD) and its variants are particularly popular across multiple disciplines; see, for example, \cite{Lu2019PODreview} for a review of POD-based methods for MOR in a variety of research areas. Additionally, we refer to \cite{BeCoOhlWill15,benner2020model} and \cite{grassle2018pod} for a comprehensive introduction to POD applied to MOR and \cite{Benner2015surveyPROM} for a survey in the context of parametrized PDEs.
In particular, POD-based reduced-order modeling has been successfully applied in engineering, in the context of fluid mechanics \citep{GIRFOGLIO2022105536,ballarin_supremizer_2015,nonino_monolithic_2021}, image processing \citep{Vaccaro1991SVD_signal}, rotor dynamics \citep{Lu2015RotorSys,Lu2016RotorSys}, for structural damage detection \citep{Eftekhar2017DamageDetection}, and in the area of structural dynamics \citep{Kerschen2005}. POD-based methods can be also used to solve optimal control problems \citep{Negri2013PODControl,Brunton2015CLturbolence,Ballarin2022spaceTimePODControl,Volkwein2001}. Moreover, POD-based ROMs can be physically interpreted based on the concept of proper orthogonal modes (POMs) \citep{Kerschen2005}. 
However, while the theory is solid for linear and some non-linear systems \citep{benner2020model}, applying ROMs to tasks such as hydrogels fabrication by printing is largely unexplored. In this work, we exploit the capabilities of POD to predict the transient diffusion-deformation of hydrogels in co-axial printing and its efficiency in tasks many-query such as material parameter identification and uncertainty propagation.  

Different POD-based algorithms have been proposed in the literature. Global POD methods focus on the estimation of global POD basis functions \citep{BeCoOhlWill15,benner2020model}. In contrast, local POD methods are based on snapshot clustering \citep{Sahyoun2013localPOD}.  
Adaptive POD methods have been developed to improve POD basis functions construction by choosing additional snapshot locations in an optimal way \citep{Lass2014adaptivePOD}.
Incremental POD has also been proposed, allowing an adaptive enrichment of the POD basis in case of unforeseen changes in the solution behavior \citep{Fareed2018iPOD,Fischer2024iPOD}.
Transient POD methods focus on snapshots of time-dependent problems \citep{Lu2015TransientPOD}. The so-called nested-POD method applies to time-dependent partial differential equations (PDEs) and enables a two-step dimensionality reduction: first in the temporal domain and then in the spatial domain, which also includes parameter information \citep{Kadeethum2021DDMOR,Kadeethum2022CAEN}. This approach has been shown to be particularly useful when the PDE requires many steps to be approximated. 
In this work, we adopt the nested-POD approach and compare it to the traditional POD.

In this paper, we consider a specialized theory for small deformations superposed on a previously homogeneously swollen gel as presented by \cite{Anand2020CMM_solids_Book}. This model is derived following rigorous thermodynamic concepts and can be considered the linearized version of the more general diffusion-deformation theory presented by \cite{Chester2010DiffDeform} and \cite{Chester2015Abaqus}.
This diffusion-deformation model was recently validated by \cite{Chen2020LinearHydrogelsExp} and showed good agreement with experimental data to study the diffusion-deformation process within the first three hours.
If long-time deformations are present, a nonlinear material model accounting for the hyperelastic deformation of the polymer network and describing the mixing of the solvent with the polymer network should be considered \citep{Chester2010DiffDeform,Bouklas2015Nonlinear,Chester2015Abaqus,Urrea2023DiffDef_hydrogels}.
We implement two POD-based ROMs, study their accuracy, and evaluate the computation speed-up to assess its suitability for many-query scenarios.

The main contributions of the paper can be summarized as follows:
\begin{enumerate}
    \item We present the first work that considers POD-based parametric ROMs for the diffusion-deformation of hydrogels.
    
    \item We demonstrate the effectiveness of the obtained parametric ROMs for material parameter identification and uncertainty propagation for hydrogel printing processes.
\end{enumerate}

The paper is organized as follows: Section~\ref{sec:gels_theory} summarizes the coupled diffusion-deformation theory of gels in the small deformation regime. 
Section~\ref{sec:MOR} introduces the POD framework. Section~\ref{sec:inverse} states the material parameter identification problem enabled by the reduced-order model. In Section~\ref{sec:simulation_results}, several simulations are conducted and computationally analyzed. Concluding remarks are given in Section~\ref{sec:conclusions}.

\section{Small deformation theory for the diffusion-deformation of hydrogels}
\label{sec:gels_theory}

Diffusion-deformation theories of gels integrate fluid transport and solid mechanics principles to explain how gels swell or shrink due to fluid movement and how these processes influence their shape and mechanical properties \citep{Chester2010DiffDeform,Bouklas2015Nonlinear,Chester2015Abaqus,Urrea2023DiffDef_hydrogels}. In a diffusion-deformation model of gels, the driving forces are fluid diffusion into or out of the gel and mechanical deformation due to external loads or internal stresses. The model should allow for predicting swelling or shrinking behaviors, stress distributions, and changes in mechanical properties over time. The primary outcomes of this model include the gel's deformation patterns and mechanical responses under varying conditions. Essential parameters for model calibration encompass the diffusion coefficient, the gel matrix's mechanical properties such as shear and bulk modulus, and fluid-gel interaction parameters such as Flory's interaction parameter.

This section summarizes the theory describing the diffusion-deformation mechanisms in elastomeric gels that undergo small deformations under isothermal conditions. 
The model is derived as a special case of a more general theory of chemoelasticity \citep{Anand2015LinearPoroelasticity,Anand2020CMM_solids_Book}.
This linear theory is suitable for capturing swelling and drying for the ﬁrst few hours of the diffusion-deformation process \citep{Anand2020CMM_solids_Book,Chen2020LinearHydrogelsExp}.
For example, the theory can describe the diffusion-deformation of hydrogels in bioprinting applications where the hydrogel is mixed in the printer's nozzle with living cells and nutrients necessary for cell growth and proliferation, or in co-axial printing where the hydrogel is directly crosslinked in the printer's nozzle \citep{Kjar2021coaxial_print}.

As notation rules, we denote gradient $\text{grad}(\bullet)$ and divergence $\text{div}(\bullet)$. The time derivative of any field is denoted by $\partial_t (\bullet)$. The operator $\text{tr}(\mathbf{A})$ refers to the trace of the second-order tensor $\mathbf{A}$.
We denote the spatial dimension with $d$, and in this section, we exclusively work with $d=3$.
Finally, let $I:=\{0,T_f\}$ be the time interval with end time value $T_f>0$.

\subsection{Kinematics of the deformation}

Consider a continuum homogeneous elastomeric body $\mathcal{B}$ within the Euclidean space $\mathbb{R}^d$ and its boundary $\partial \mathcal{B} = \partial \mathcal{B}{\mathbf{u}} \cup \partial \mathcal{B}{\bar{\mathbf{t}}} = \partial \mathcal{B}_{R}$. Here, $\partial \mathcal{B}{\mathbf{u}}$ denotes the displacement (Dirichlet) boundary, $\partial \mathcal{B}{\bar{\mathbf{t}}}$ the traction (Neumann) boundary, and $\partial \mathcal{B}_{R}$ the fluid flux (Robin) boundary. The outward normal vector to the domain boundaries is denoted by $\mathbf{n}\in\mathbb{R}^d$.
The displacement field is defined as $\mathbf{u}:\mathcal{B}\to\mathbb{R}^d$ and the deformation is described by the displacement gradient tensor, $\mathbf{H}:\mathcal{B}\to\mathbb{R}^{d \times d}$,
\begin{equation}\label{eq:H_grad_u}
\mathbf{H} = \text{grad} (\mathbf{u}).
\end{equation}

For small deformations, the strain tensor can be approximated by the linearized strain tensor, $\bm{\varepsilon}:\mathcal{B}\to\mathbb{R}^{d \times d}$,
\begin{equation}\label{eq:strain}
\bm{\varepsilon} = \frac{1}{2} (\mathbf{H} + \mathbf{H}^T),
\end{equation}
which represents the symmetric part of the displacement gradient.

\subsection{Governing partial differential equations}
The two governing PDEs for the quasi-static vector-valued displacements $\mathbf{u}:\mathcal{B}\to\mathbb{R}^d$ and scalar-valued, time-dependent, fluid content in terms of the concentration $c:\mathcal{B}\times I\to\mathbb{R}$ consist of:

\begin{enumerate}
    \item The local form of the \textbf{balance of linear momentum}:
    \begin{equation}\label{eq:linear_momentum}
        \begin{cases}
            \text{div} (\bm{\sigma}) + \mathbf{b} = \bm{0}, & ~ \text{in} ~ \mathcal{B}, \\
            \mathbf{u} = \bar{\mathbf{u}}, & ~ \text{on} ~ \partial \mathcal{B}_{\mathbf{u}}, \\
            \bm{\sigma} \cdot \mathbf{n} = \bar{\mathbf{t}}, & ~ \text{on} ~ \partial \mathcal{B}_{\bar{\mathbf{t}}}, \\
            \bm{\sigma}\vert_{t = 0} = \bm{\sigma}_0, & ~ \text{in} ~ \mathcal{B}\, .
        \end{cases}
    \end{equation}

    Here, $\bm{\sigma}:\mathcal{B}\to\mathbb{R}^{d \times d}$ denotes the Cauchy stress tensor, specified below, and $\bm{\sigma}_0:\mathcal{B}\to\mathbb{R}^{d \times d}$ its initial value at $t = 0$. 
    External actions consist of body forces per unit deformed volume $\mathbf{b}:\mathcal{B}\to\mathbb{R}^d$. Moreover, boundary conditions prescribe displacements $\bar{\mathbf{u}}:\mathcal{B}\to\mathbb{R}^d$ and traction $\bar{\mathbf{t}}:\partial\mathcal{B}_{\bar{\mathbf{t}}}\to\mathbb{R}^{d}$ on separate portions of the boundary. Notice that inertial effects have been neglected due to the considerably slow dynamics of the fluid diffusion evolution w.r.t. the time scale of the wave propagation. 

    \item The local form of the \textbf{mass balance of fluid content} inside the hydrogel:
    \begin{equation}\label{eq:fluid_balance_robin}
        \begin{cases}
        \partial_t c + \text{div} ( \mathbf{j} ) = 0, & ~ \text{in} ~ \mathcal{B} \times I,\\
        -\mathbf{j} \cdot \mathbf{n} = \alpha_R (\mu - \mu_{\infty}), & ~ \text{on} ~ \partial \mathcal{B}_{R}\times I,\\
        \mu\vert_{t = 0} = \mu_0, & ~ \text{in} ~ \mathcal{B}\times \lbrace t=0 \rbrace.
        \end{cases}
    \end{equation}

    Boundary conditions are defined by a Robin condition on the boundary $\partial \mathcal{B}_{R}$, which prescribes a relationship between the fluid flux $\mathbf{j}$ and the chemical potential $\mu$. Specifically, $\alpha_R$ is a proportionality constant, $\mu$ is the chemical potential at the boundary, and $\mu_{\infty}:\partial \mathcal{B}_{R}\times I\to\mathbb{R}$ represents a reference chemical potential. This condition models the proportional flux response to the difference in chemical potential at the boundary and is considered to be more consistent with experimental observations of fluid absorption in hydrogels \citep{Chen2020LinearHydrogelsExp}.
    Furthermore, $\mu_0: \mathcal{B}\times \lbrace t=0 \rbrace  \to \mathbb{R}$ refers to the initial value of the chemical potential inside the hydrogel. Notice that the mass balance of fluid content is written in terms of $c$ and $\mathbf{j}$, but its corresponding boundary and initial conditions involve the chemical potential $\mu$. The connection of equation \eqref{eq:fluid_balance_robin} with $\mu$ becomes clear by introducing a constitutive relation for $\mathbf{j}$, for example, through Fick's laws of diffusion. 
\end{enumerate}

\subsection{Thermodynamically consistent constitutive theory for linear isotropic gels}

In this subsection, we present a thermodynamically consistent constitutive theory describing the coupled diffusion-deformation mechanisms for linear isotropic gels. 
Only the main ingredients are provided, but the step-by-step derivation of the constitutive model is detailed in \ref{app:const_theory}.

For linear isotropic hydrogels, the following thermodynamically-consistent constitutive relation can be established for \textbf{Cauchy's stress} 
\begin{equation}\label{eq:specific_sigma}
    \bm{\sigma}(\mu, \bm{\varepsilon}) = 2 G  \bm{\varepsilon} + \lambda^d (\text{tr} \bm{\varepsilon}) \bm{1} + \frac{\chi}{\Lambda}(\mu - \mu_{0}) \bm{1},
\end{equation}
with $G$ representing the elastic shear modulus, $\lambda^{d}$ the drain Lamé parameter, $\bm{1}$ the identity tensor, $\chi$ the stress-chemical modulus, and $\Lambda$ the chemical modulus.

Additionally,  the following constitutive relation can be established for \textbf{the species concentration} as a function of the chemical potential
\begin{equation}\label{eq:specific_mu}
    c(\mu, \bm{\varepsilon}) = c_{0} - \frac{\chi}{\Lambda} \text{tr} \bm{\varepsilon} - \frac{1}{\Lambda}(\mu - \mu_{0}).
\end{equation}

Moreover, we append a thermodynamically motivated choice for the fluid flux $\mathbf{j}$ based on Fick's law for species diffusion. That is, $\mathbf{j}$ is proportional to the gradient of the chemical potential, namely,
\begin{equation}\label{eq:flux_j}
    \mathbf{j} = - \mathbf{M} ~ \text{grad} (\mu),
\end{equation}
with $\mathbf{M}$ as the species mobility tensor defined as
\begin{equation}\label{eq:mobility_isotropic}
    \mathbf{M} = \left(\frac{D}{k_B \bar{T}}\right) \bm{1},
\end{equation}
where $D$ is a constant for fluid diffusivity, $k_B$ is the Boltzmann constant, and $\bar{T}$ is the absolute temperature.

The final model is characterized by the four independent material parameters, two from isotropic linear elasticity $[G, {\lambda}^{d}]$ and the other two from the fluid equations $[\Omega, D]$, where $\Omega$ is the molar volume of the fluid as defined in \ref{app:const_theory}.

It is worth mentioning that the theory introduced here can be linked with the classical theories of chemoelasticity and linear poroelasticity; see \cite{Anand2015LinearPoroelasticity} and \cite{Anand2020CMM_solids_Book} [Chapters 15 and 16] for more details. The related transformations are explained in Appendix~\ref{app:const_theory}. 

\subsection{Coupled diffusion-deformation model in normalized form}

To facilitate the analysis of the model derived in \ref{app:const_theory}, Problem~\ref{eq:full_model}, the physical quantities are normalized as follows
\begin{equation} \label{eq:normalized_parm}
\begin{array}{cc}
    \bm{X} \rightarrow \frac{\bm{x}}{l}, & 
    T \rightarrow \frac{t D}{l^{2}},  \\
    \mu^* \rightarrow \frac{{\mu}}{k_{B}\bar{T}}, &
    \bm{\sigma}^* \rightarrow \frac{\bm{\sigma}}{G}, \\
    \mathbf{b}^* \rightarrow \frac{\bm{b}l}{G}, &
    \bm{t}^* \rightarrow \frac{\bm{t}}{G}, \\
    \lambda^* \rightarrow \frac{\lambda^{d}}{G}, &
    A \rightarrow \frac{k_{B}\bar{T}}{G\Omega}.
\end{array}
\end{equation}

In equation~\eqref{eq:normalized_parm}, $l$ is the reference length of the hydrogel, $\lambda^*$ is the dimensionless first Lamé constant, which indicates how much volumetric versus shear deformation contributes to the material response, and $A$ is the dimensionless scaling factor for the chemical potential's influence on the stress. This normalization closely follows the methodology presented by \cite{Chen2020LinearHydrogelsExp}. 

The reader should note that the normalized model only contains two non-dimensional parameters, compared to six parameters in the original formulation.
Moreover, as the time-dependent chemical potential enters into the displacements, the governing PDEs become quasi-static, indirectly depending also on time. Thus, we write $\mathbf{u}: \mathcal{B} \times I \to \mathbb{R}^d$ in the remainder of the paper.

The resulting coupled PDE model with normalized quantities reads:
\begin{problem}[Strong form linear diffusion-deformation of gels]
    Given $\bar{\mathbf{u}},\bar{\mathbf{t}}^*,\mu^*_{\infty}$ as boundary data and $\bm{\sigma}^*_0,\mu^*_0$ as initial data, find 
    $\bm{u}:\mathcal{B}\times I\to\mathbb{R}^d$ and $\mu^*:\mathcal{B}\times I\to\mathbb{R}$ such that
\begin{align}\label{eq:full_model_dimensionless}
    \text{div} (\bm{\sigma}^*) + \mathbf{b}^* = \bm{0}, & ~ \text{in} ~ \mathcal{B} \times I,\\
    \text{tr} \left( \partial_t \bm{\varepsilon} \right) = \text{div} \left( \text{grad} \left( \mu^* \right)\right), & ~ \text{in} ~ \mathcal{B} \times I,\\
    \mathbf{u} = \bar{\mathbf{u}}, & ~ \text{on} ~ \partial \mathcal{B}_{\mathbf{u}} \times I, \\
            \bm{\sigma}^* \cdot \mathbf{n} = \bar{\mathbf{t}}^*, & ~ \text{on} ~ \partial \mathcal{B}_{\bar{\mathbf{t}}} \times I, \\
            \bm{\sigma}^*\vert_{t = 0} = \bm{\sigma}^*_0, & ~ \text{in} ~ \mathcal{B} \times \lbrace T = 0 \rbrace , \\
          \text{grad} (\mu^*) \cdot \mathbf{n} = \alpha_R (\mu^* - \mu^*_{\infty}), & ~ \text{on} ~ \partial \mathcal{B}_{R}\times I, \label{eq:robin_BC_final} \\
        \mu^*\vert_{t = 0} = \mu^*_0, & ~ \text{in} ~ \mathcal{B}\times \lbrace T=0 \rbrace, 
\end{align}
with the constitutive law for stresses in normalized form given by
\begin{equation}\label{eq:dimensionless_sigma}
    \bm{\sigma}^* = 2 \bm{\varepsilon} + \lambda^* (\text{tr} \bm{\varepsilon}) \bm{1} -A(\mu^* - \mu^*_{0}) \bm{1},
\end{equation}
and the strain tensor defined as in equation \eqref{eq:strain}, which now contains only two dimensionless material parameters, namely $[\lambda^*, A]$, defined in equation \eqref{eq:normalized_parm}. 
\end{problem}

The reader should notice that, for simplicity in the notation, the asterisks in the normalized quantities will be removed in the subsequent sections.

\subsection{Weak formulation}

The solution of the coupled PDE system given by equations \eqref{eq:full_model_dimensionless} - \eqref{eq:dimensionless_sigma} consists of a vector-valued field of displacements $\mathbf{u}$ and a scalar-valued field of the chemical potential $\mu$. 

In the following, we derive the weak formulation. We adopt standard notation for the usual Lebesgue and Sobolev spaces, see, for example, \cite{Wlo87}. The functional space $H^1(\mathcal{B})^d$ is a Sobolev space that consists of functions defined on a bounded domain $\mathcal{B} \subset \mathbb{R}^d$, with square integrable partial derivatives up to the first order. To this end, we define the trial and test spaces as follows 
\begin{align*}
    Q &:= H^1 \left( \mathcal{B} \right), \qquad\qquad 
    V := \left( H^1 \left( \mathcal{B} \right) \right)^d, \\ 
    V_{\bar{\mathbf{u}}} &:= \Big\lbrace \mathbf{u} \in V \Big\vert ~ \mathbf{u} = \bar{\mathbf{u}} ~~ \text{on} ~~ \partial \mathcal{B}_{\mathbf{u}} \Big\rbrace, 
    \quad V_{\bm{0}} : = \Big\lbrace \mathbf{u} \in V \Big\vert ~ \mathbf{u} = \bm{0} ~~ \text{on} ~~ \partial \mathcal{B}_{\mathbf{u}} \Big\rbrace.
\end{align*}
To formulate both problem statements in an abstract fashion, we introduce 
for the displacement system the bilinear form $a((\mu, \mathbf{u}))(\mathbf{v})$. Furthermore, 
let $b(\mathbf{v})$ be the given right-hand side data. Next, for the balance of fluid concentration, we use $c((\mu, \mathbf{u}))(q) $ and $d(q)$.
Then, the weak formulation can be written as
\begin{problem}[Weak form] 
\label{form_1}
    Find $(\mathbf{u},\mu) \in V_{\bar{\mathbf{u} }} \times Q$, with $\mu(0) = \mu_0$ and $\sigma(0) = \sigma_0$, 
    such that for $t\in I$ it holds
    \begin{equation}\label{eq:weak_form}
    \bm{\mathcal{M}}( (\mu, \mathbf{u}), t) :=
        \begin{cases}
            a((\mu, \mathbf{u}))(\mathbf{v}) + b(\mathbf{v}) & = \mathbf{0}, ~~ \forall \mathbf{v} 
            \in V_0, \\
            c((\mu, \mathbf{u}))(q) + d(q) & = 0, ~~ \forall q \in Q,
        \end{cases}
    \end{equation}
    where
    \begin{align}
        a((\mu, \mathbf{u}))(\mathbf{v}) & = \int_{\mathcal{B}} \bm{\sigma} ( \mu, \mathbf{u}): \text{grad} (\mathbf{v}) dV, \\
        b(\mathbf{v}) & = - \int_{\mathcal{B}} \mathbf{b} \cdot \mathbf{v} dV, \\
        c((\mu, \mathbf{u}))(q) & = \int_{\mathcal{B}} \text{tr} \left( \partial_t \bm{\varepsilon}(\mathbf{u}) \right) \cdot q dV + \int_{\mathcal{B}} \text{grad} (\mu) \cdot \text{grad} (q) dV - \int_{\partial \mathcal{B}_R} \alpha_R \mu\cdot q dS, \\
        d(q) & = - \int_{\partial \mathcal{B}_R} \alpha_R \mu_{\infty}\cdot q dS,
    \end{align}
    with $``~:~"$ denoting the double contraction of the second-order tensors $\bm{\sigma}$ and $\text{grad} (\mathbf{v})$, where $\bm{\sigma} ( \mu, \mathbf{u})$ is defined by equation \eqref{eq:dimensionless_sigma}, which couples the two balance equations.
\end{problem}


\subsection{Discrete weak formulation and numerical solution}

In order to deal with efficient POD-based ROMs, we are required to fulfill the assumption of affine parametric dependence on the operators appearing in the weak form \eqref{eq:weak_form} at the FOM level. This allows us to decouple the construction stage of the reduced-order space (offline) from the parametric evaluation stage (online).

To meet this requirement, let us assume that the computational domain $\mathcal{B}$ is decomposed into triangular elements. We employ Taylor-Hood elements such that the inf-sup condition is fulfilled, that is,
$V_{h, {\mathbf{0}}} \times Q_h \subset V_{{\mathbf{0}}} \times Q$, with quadratic functions in $V_{h}$ and linear functions in $Q_h$. Then, the parametric semi-discrete variational monolithic formulation for the diffusion-deformation model reads:
\begin{problem}[Parametric spatial semi-discrete weak form]
\label{param_disc_form}
    Given the material parameter $\bm{\theta}=[\lambda, A]\in \mathbb{R}^{d_{\bm{\theta}}}$ and model coefficients $\bm{\Theta}:= [2, \bm{\theta}] \in \mathbb{R}^{d_{\bm{\theta}}+1}$, find $(\mathbf{u}_h, \mu_h) \in \{\bar{\mathbf{u}}_h|_{\partial \mathcal{B}_{\bar u}} + V_{h,{\mathbf{0}}}\} \times Q_h$ such that for $t\in I$ it holds

    \begin{equation}\label{eq:disc_weak_form}
    \bm{\mathcal{M}}( (\mu_h, \mathbf{u}_h), t; \bm{\theta} ) :=
    \begin{cases}
    \bm{\Theta}~a((\mu_h, \mathbf{u}_h)) (\mathbf{v}_h) = b (\mathbf{v}_h), \quad \forall \mathbf{v}_h \in V_{h, \mathbf{0}}, \\
    c ((\mu_h, \mathbf{u}_h))(q_h) = d (q_h), \quad \forall q_h \in Q_{h},
    \end{cases}
    \end{equation}

    where equation~\eqref{eq:disc_weak_form}$_1$ can be reformulated as
    \begin{equation}
        \bm{\Theta}~a((\mu_h, \mathbf{u}_h)) (\mathbf{v}_h) = \sum_{m=1}^{3} \Theta_m a_m (\mathbf{u}_h, \mathbf{v}_h),
    \end{equation}
    
    with
    \begin{equation}
    \begin{aligned}
        a_1 (\mathbf{u}_h) (\mathbf{v}_h) &= \int_{\mathcal{B}} \bm{\varepsilon} (\mathbf{u}_h) : \text{grad} (\mathbf{v}_h) \, dV, \\
        a_2 (\mathbf{u}_h) (\mathbf{v}_h) &= \int_{\mathcal{B}} (\text{tr} \bm{\varepsilon} (\mathbf{u}_h)) \bm{1} : \text{grad} (\mathbf{v}_h) \, dV, \\
        a_3 ((\mu_h, \mathbf{u}_h)) (\mathbf{v}_h) &= -\int_{\mathcal{B}} (\mu_h - \mu_0) \bm{1} : \text{grad} (\mathbf{v}_h) \, dV,
    \end{aligned}
    \end{equation}
    
    and
    \begin{equation}
        b (\mathbf{v}_h) = - \int_{\mathcal{B}} \mathbf{b} \cdot \mathbf{v}_h \, dV .
    \end{equation}
    
    Additionally, equation~\eqref{eq:disc_weak_form}$_2$ can be written analogously as
    
    \begin{equation}
        c((\mu_h, \mathbf{u}_h)) (q_h) = \sum_{n=1}^{3} c_n ((\mu_h, \mathbf{u}_h), q_h)
    \end{equation}
    
    with
    \begin{equation}\label{eq:mu_disc_bilinear}
    \begin{aligned}
        c_1 (\mathbf{u}_h) (q_h) &= \int_{\mathcal{B}} (\text{tr} (\partial_t \bm{\varepsilon} (\mathbf{u}_h))) q_h \, dV, \\
        c_2 (\mu_h)(q_h) &= \int_{\mathcal{B}} \text{grad} (\mu_h) \cdot \text{grad} (q_h) \, dV, \\
        c_3 (\mu_h)(q_h) &= \int_{\partial \mathcal{B}_{R}} \alpha_R (\mu_h - \mu_{\infty}) q_h \, dS,
    \end{aligned}
    \end{equation}
    
    and
    \begin{equation}
        d (q_h) = - \int_{\partial \mathcal{B}_{R}} \alpha_R \mu_{\infty} q_h \, dS.
    \end{equation}
    Here, $\bm{\mathcal{M}}( (\mu_h, \mathbf{u}_h), t; \bm{\theta} )$ is the semi-discrete operator that captures the transient behavior of the coupled diffusion-deformation problem and is affine parameterized by the material parameters $\bm{\theta}=[\lambda, A]$. 
\end{problem}

The time-dependent operator defined in equation \eqref{eq:mu_disc_bilinear}$_1$ is further discretized in time with the first-order A-stable implicit Euler scheme.
The arising linear system of equations in Problem~\ref{param_disc_form} is solved with a sparse direct solver. 
The implementation is done in FEniCS. All details of this FOM realization can be found in \cite{Urrea2023DiffDef_hydrogels}.

\section{Parametric POD-based reduced order modeling}
\label{sec:MOR}

In this section, we describe parametric POD-based and parametric nested-POD approaches as reduced-order modeling strategies applied to the transient diffusion-deformation of hydrogels. Such a ROM construction allows to speed up the solution of the coupled model presented in Section~\ref{sec:gels_theory}, which enables its use in many-query scenarios.

\begin{figure}[!htb]
\center{\includegraphics[width=1.0\textwidth]{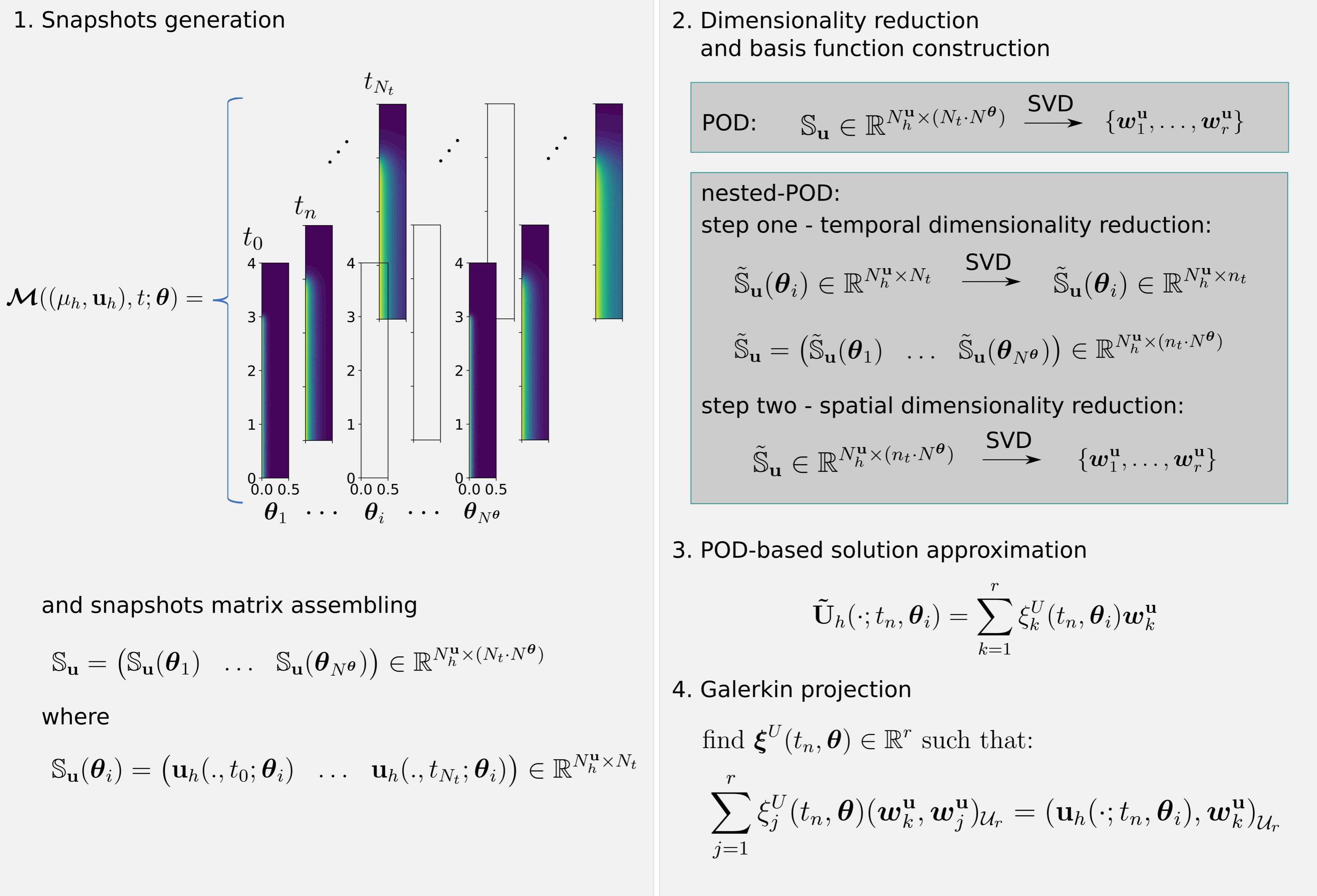}}
\caption{\textbf{Illustration of the offline phase workflow for the POD-based ROM's construction.}}
\label{fig:POD_workflow_illustration}
\end{figure}

The computation in the reduced-order modeling strategy is divided into two phases: offline and online. 
\begin{enumerate}[label=(\arabic*)]
    \item \textbf{The offline phase} refers to the parametric ROM construction. It mainly consists of two steps: 
    \begin{enumerate}[label=(1\alph*)]
        \item Collect a set of snapshots that are FOM solutions for different parameter values and time instances. The simulation time interval for each parameter instance remains the same.
        \item Find an optimal reduced-order approximation of the primary variables $(\mu_h, \mathbf{u}_h)$ by projecting the snapshot matrix into a reduced linear subspace using Galerkin approach.
    \end{enumerate}
    Figure~\ref{fig:POD_workflow_illustration} illustrate the main steps of the offline phase for the parametric ROM construction.
    \item \textbf{The online phase} refers to the use of the ROM as a surrogate of the FOM to predict the diffusion-deformation of hydrogels in many-query tasks. Here, the parametric ROM is used for parameter identification; see Section~\ref{sec:simulation_results}.
\end{enumerate}
The offline steps are explained in detail below. 
The starting point is the assembly of the snapshot matrix in Section~\ref{sec:partioning_primary}, followed by the singular value decomposition (SVD) in Section~\ref{sec:pod}. We introduce the nested-POD approach as a computationally efficient variant in Section~\ref{sec:nested_pod}. The Galerkin projection is outlined in Section~\ref{sec:galerkin_projection}. The implementation of the parametric POD and nested-POD-based ROMs is inspired by \cite{Lassila2014ROM} and \cite{Kadeethum2021DDMOR}.

\subsection{Assembly of snapshots matrix}\label{sec:partioning_primary}
The snapshots are obtained by solving the model
\begin{equation}
    \bm{\mathcal{M}}((\mu_h, \mathbf{u}_h), t; \bm{\theta}) = \bm{0}.
\end{equation}
This results in a global snapshot matrix $\mathbb{S}\in \mathbb{R}^{(N_h^{\mu}+N_h^{\mathbf{u}}) \times (N_t \cdot N^{\bm{\theta}})}$, with
\begin{enumerate}[label=\textbf{\roman*.}]
    \item $N^{\mathbf{u}}_h$ the number of displacement degrees of freedom,
    \item $N^{\mu}_h$ the number of chemical potential degrees of freedom,
    \item $N_t$ the number of time steps, and
    \item $N^{\bm{\theta}}$ the number of parameter instances used for training. 
\end{enumerate}

Different assembly strategies arise from splitting the subset global snapshots into subsets in order to enhance computational efficiency. Here, we consider partitioning in the primary variables $(\mu_h, \mathbf{u}_h)$ as well as in space and time. The reader should note that the latter approach is also referred to as the nested-POD approach.

\textbf{Partitioning of primary variables:} To this end, we consider a partitioned approach to derive reduced bases for $\mu_h$ and $\mathbf{u}_h$ separately. We focus on the displacement field in what follows, although a similar procedure applies to the chemical potential. 
Given  the parameter instances $\bm{\theta}_i$, $\forall i \in N^{\bm{\theta}}$ of the training dataset,  the corresponding \textbf{temporal snapshot matrix} $\mathbb{S}_{\mathbf{u}}(\bm{\theta}_i)\in \mathbb{R}^{N^{\mathbf{u}}_{h} \times N_t}$ is assembled as:
\begin{equation}\label{eq:snapshots_time}
\mathbb{S}_{\mathbf{u}}(\bm{\theta}_i) = 
\begin{pmatrix}
    \mathbf{u}_h (.,t_0;\bm{\theta}_i) & \hdots &\mathbf{u}_h (.,t_{N_t};\bm{\theta}_i)
\end{pmatrix} 
\end{equation}
for the displacement field $\mathbf{u}_h(.,t_n;\bm{\theta}_i)$ at a specific time $t_n$, for $n = 0, \dots, N_t$, and parameter instance $\bm{\theta}_i$.

The \textbf{complete snapshots matrix} $\mathbb{S}_{\mathbf{u}}\in \mathbb{R}^{N^{\mathbf{u}}_{h} \times (N_t \cdot {N^{\bm{\theta}}})},$ is then constructed by aggregating the individual temporal snapshot matrices $\mathbb{S}_{\mathbf{u}}(\bm{\theta}_i)$ as:
\begin{equation}\label{eq:snapshots_space_time}
    \mathbb{S}_{\mathbf{u}} = 
    \begin{pmatrix}
        \mathbb{S}_{\mathbf{u}}(\bm{\theta}_1) & \hdots & \mathbb{S}_{\mathbf{u}}(\bm{\theta}_{N^{\bm{\theta}}})
    \end{pmatrix}.
\end{equation}

Consequently, each row in $\mathbb{S}_{\mathbf{u}}$ corresponds to a degree of freedom in the spatial discretization, and each column corresponds to the state of the system at a particular time for a particular parameter sample. 

\subsection{POD for the displacement field}\label{sec:pod}
In this subsection, we describe the key parts of the POD for the displacement field. The chemical potential POD is computed in an analogous fashion.
In synthesis, POD reduces the dimensionality of $\mathbb{S}_{\mathbf{u}}$ \eqref{eq:snapshots_space_time} by computing a matrix $\mathbb{V} \in \mathbb{R}^{N_h^{\mathbf{u}} \times r}$, which represents the best approximation in the least-squares sense to the snapshot matrix $\mathbb{S}_{\mathbf{u}}$. Here, $r \ll N_t \cdot N^{\bm{\theta}}$ is the number of retained modes based on analyzing the singular values that retain the most significant information in the snapshots matrix.
We refer the reader to equation \eqref{eq:retained_ener}.

The POD solution is then derived by solving the optimization problem:
\begin{align}\label{eq:min_eigen}
\mathbb{V}^{*} = & \operatorname{arg\, min}_{\mathbb{V} \in \mathbb{R}^{N_h^{\mathbf{u}} \times r}} \lVert \mathbb{S}_{\mathbf{u}} - \mathbb{V}\mathbb{V}^T \mathbb{S}_{\mathbf{u}} \rVert_F^2 = \sum_{i = r + 1}^{(N_t \cdot N^{\bm{\theta}})} \sigma_i, 
\end{align}
where $\lVert \cdot \rVert_F$ denotes the Frobenius norm, which measures the distance between $\mathbb{S}_{\mathbf{u}}$ and its approximation $\mathbb{V}\mathbb{V}^T \mathbb{S}_{\mathbf{u}}$ in terms of all entries of the snapshots matrix. Moreover, $\sigma_i$ refers to the $i$-th singular value of $\mathbb{S}_{\mathbf{u}}$ and $\mathbb{V}\mathbb{V}^T$ yield the identity matrix with dimension of $N_h^{\mathbf{u}} \times N_h^{\mathbf{u}}$.

The singular values $\sigma_i$ provide quantitative guidance for choosing the size of the POD basis. That is, the number of modes $r$ required to capture a desired fraction of energy of the FOM, measured by the fraction of retained energy
\begin{equation}\label{eq:retained_ener}
\mathcal{E} = \frac{E_r}{E_{\text{total}}} = \frac{\sum_{i=1}^{r} \sigma_i^2}{\sum_{i=1}^{(N_t \cdot N^{\bm{\theta}})} \sigma_i^2} \geq \eta,
\end{equation}
where $E_{\text{total}}$ is the total energy of the system and $E_r$ is the retained energy by \(r\) modes. Consequently, the number of modes $r$ is chosen such that the retained energy ratio meets a predefined target fraction \(\mathcal{E} \geq \eta \in [0, 1]\).

\subsection{Nested-POD approach for the displacement field}\label{sec:nested_pod}
For a large number of parameters $N^{\bm{\theta}}$ and/or time steps $N_t$, applying SVD on the snapshot matrix \eqref{eq:snapshots_space_time} may become inefficient, requiring large memory storage. An alternative is to apply the nested-POD approach; see, for instance, \cite{Kadeethum2021DDMOR}, and also our own study in Figure~\ref{fig:2D_square_ROM_param_sampling_wall_time} (Section \ref{sec:simulation_results}).

The snapshot matrix $\mathbb{S}_{\mathbf{u}}$ \eqref{eq:snapshots_space_time} can be further decomposed into temporal and space parts to apply the nested-POD approach.
That is, the time dimension is reduced first, followed by the spatial dimension for the parameter samples. 
The dimensionality reduction steps can be summarized as follows:
\begin{enumerate}
    \item \textbf{Temporal dimensionality reduction:} POD is applied individually for  every parameter sample $\bm{\theta}_i$ on the snapshots matrix $\mathbb{S}_{\mathbf{u}}(\bm{\theta}_i)$ \eqref{eq:snapshots_time}.
    In this way, we obtain the basis functions for the first $n_t \ll N_t$ temporal modes and for each $\bm{\theta}_i$. The snapshot matrix after the temporal reduction is represented as $ \Tilde{\mathbb{S}}_{\mathbf{u}}(\bm{\theta}_i) \in \mathbb{R}^{N^{\mathbf{u}}_h \times n_t}$:
    \begin{equation}
        \Tilde{\mathbb{S}}_{\mathbf{u}}(\bm{\theta}_i) =
        \begin{pmatrix}
            \tilde{\bm{w}}^{\mathbf{u}}_1 & \dots & \tilde{\bm{w}}^{\mathbf{u}}_{n_t}
        \end{pmatrix},
    \end{equation}
    with $\tilde{\bm{w}}^{\mathbf{u}}_i$ referring to the basis functions in the temporal domain.
    
    \item  \textbf{Spatial dimensionality reduction:} after the temporal dimension has been reduced for each parameter sample $\bm{\theta}_i$, we assemble a new snapshot matrix $\Tilde{\mathbb{S}}_{\mathbf{u}} \in \mathbb{R}^{N^{\mathbf{u}}_{h} \times (n_t \cdot N^{\bm{\theta}})}$ as
    \begin{equation} \label{eq:compressed_time_disp}
    \Tilde{\mathbb{S}}_{\mathbf{u}} = 
    \begin{pmatrix}
        \Tilde{\mathbb{S}}_{\mathbf{u}}(\bm{\theta}_1) & \hdots & \Tilde{\mathbb{S}}_{\mathbf{u}}(\bm{\theta}_{N^{\bm{\theta}}})
    \end{pmatrix}.
    \end{equation}
    Then, in analogy to the optimization problem \eqref{eq:min_eigen}, POD is applied again on $\Tilde{\mathbb{S}}_{\mathbf{u}}$ to retain the $r$ first modes and construct spatial bases functions.
\end{enumerate}
For both the temporal and spatial dimensionality reduction, the criterion \eqref{eq:retained_ener} is employed to select the number of required basis functions.

It is worth remarking here that the SVD algorithm is at the core of dimensionality reduction. 
In this work, we adopt the SVD algorithm provided by the RBniCS library. 
Other methods can also be used to improve the SVD efficiency further. For example, the method of snapshots \citep{Sirovich1987SnapshotsSVD} and the streaming method of snapshots \citep{Hemati2014StreamingSVD}, or modern variants such as randomized \citep{Martinsson2011RandomizedSVD} and incremental \citep{Brand2006IncrementalSVD,Kuhl2024IncrementalSVD} SVD.

\subsection{Galerkin projections}\label{sec:galerkin_projection}
In this subsection, we finally construct the reduced-order solutions using Galerkin projections.
In both the POD and nested-POD cases, only the first $r$ modes are retained during the dimensionality reduction and are employed to construct the basis functions for the reduced order space $\mathcal{U}_r$ for the displacement field $\bm{u}$. 
Thus, we map the temporal and spatial dimensions to their representation in a linear manifold, producing the best approximation in the reduced space $\mathcal{U}_r$ for $\mathbf{u}(\cdot; t, \bm{\theta})$.

The optimal representation is obtained on a reduced basis space by means of Galerkin projection. Let $(\bm{w}^{\mathbf{u}}_1, \dots, \bm{w}^{\mathbf{u}}_r)$ denote the basis functions spanning $\mathcal{U}_r$. Given a time $t_n$ and a parameter instance $\bm{\theta}_i$ in the training set, the best approximation $\mathbf{\Tilde{U}}_h (\cdot; t_n, \bm{\theta}_i)$ to $\mathbf{u}_h (\cdot; t_n, \bm{\theta}_i)$ in $\mathcal{U}_r$ is given by
\begin{equation}
    \mathbf{\Tilde{U}}_h (\cdot; t_n, \bm{\theta}_i) = \sum_{k=1}^r \xi_k^U (t_n, \bm{\theta}_i) \bm{w}^{\mathbf{u}}_k,
\end{equation}
where the coefficients $\xi_k^U$ are solutions to the Galerkin projection problem. Specifically, for given $\mathbf{u}_h (\cdot; t_n, \bm{\theta}_i)$, find $\bm{\xi}^U (t_n, \bm{\theta}) = \lbrace \xi_1^U (t_n, \bm{\theta}_1), \dots , \xi_r^U (t_n, \bm{\theta}_r) \rbrace \in \mathbb{R}^r$ such that,
\begin{equation}
    \sum_{j=1}^r \xi_j^U (t_n, \bm{\theta}) (\bm{w}^{\mathbf{u}}_k, \bm{w}^{\mathbf{u}}_j)_{\mathcal{U}_r} = \left( \mathbf{u}_h (\cdot; t_n, \bm{\theta}_i), \bm{w}^{\mathbf{u}}_k \right)_{\mathcal{U}_r}.
\end{equation}
The operator $(\bm{w}^{\mathbf{u}}_k, \bm{w}^{\mathbf{u}}_j)_{\mathcal{U}_r}$ forms a matrix that can be precomputed and stored, facilitating the efficient computation of the coefficients $\xi_k^U$ for each new instance of the parameters or time.

The reduced basis approximation $\mathbf{\hat{U}}_h (\cdot; t_n, \bm{\theta}_i)$ is computed as:
\begin{equation}
    \mathbf{\hat{U}}_h (\cdot; t_n, \bm{\theta}_i) = \sum_{k=1}^r \xi_k^U (t_n, \bm{\theta}_i) \bm{w}^{\mathbf{u}}_k.
\end{equation}

In the same fashion, we obtain the reduced basis 
approximation for the chemical potential:
\begin{equation}
    \hat{\Gamma}_h (\cdot; t_n, \bm{\theta}_i) = \sum_{k=1}^r \xi_k^{\Gamma} (t_n, \bm{\theta}_i) \bm{z}^{\mu}_k.
\end{equation}
Here, $\bm{z}^{\mu}_1, \dots, \bm{z}^{\mu}_r$ are the basis functions spanning the reduced order space $\Gamma_r$. Furthermore, $\xi_k^{\Gamma}$ are the coefficients resulting from the Galerkin projection.

We can numerically approximate the parametric discrete FOM $\bm{\mathcal{M}}( (\mu_h, \mathbf{u}_h), t; \bm{\theta} )$ in Problem~\ref{param_disc_form} by the POD-based ROM $\bm{\hat\mathcal{M}}( (\hat{\Gamma}_h, \mathbf{\hat{U}}_h), t; \bm{\theta} )$  as
\begin{equation}\label{eq:FOMbyROM}
    \bm{\mathcal{M}}( (\mu_h, \mathbf{u}_h), t; \bm{\theta} ) \approx \bm{\hat\mathcal{M}}( (\hat{\Gamma}_h, \mathbf{\hat{U}}_h), t; \bm{\theta} ),
\end{equation}
to approximate $\mu_h$ and $\mathbf{u}_h$ in the online phase.

The reader should notice that, in the online phase, the ROM should be only used for predictions of $\mu_h$ and $\mathbf{u}_h$ within the parameter range for which it has been trained. Particularly, the ROM model in equation \eqref{eq:FOMbyROM} needs to be evaluated at the new desired $\bm{\theta}$ values. If, in the online phase, it becomes apparent that predictions outside this parameter range are necessary, additional training is recommended, as extrapolation is prone to errors. 
Then, one can look into various methods proposed, for instance, by \cite{Nils2024}, \cite{Kevin2019}, or \cite{Fischer2024iPOD}. Furthermore, some research has also been focusing on using machine learning combined with ROM to tackle this issue, as done by \cite{Bernard2022}.
However, in the present study, we did not consider such cases.

Finally, we utilize the $L^1, L^2$ and $L^{\infty}$ norms to evaluate the approximation errors, for example, $\mathbf{e}_{\mathbf{u}}:= \mathbf{u}_{FOM} - \mathbf{u}_{ROM}$ between the FOM $\mathbf{u}_{h_{FOM}}$ and ROM $\mathbf{u}_{h_{ROM}}$ solutions.

\section{Material parameter identification from full-field data}\label{sec:inverse}

We exemplary consider material parameter identification from full-field observations to evaluate the ROM's performance in a many-query scenario.
A recent review on parameter identification from full-field data in computational mechanics can be found in \cite{romer2024reduced}. Here, our objective is to identify the optimal set of dimensionless material parameters $\boldsymbol{\theta}^{\text{opt}} = [\lambda^{\text{opt}}, A^{\text{opt}} ] \in \mathbb{R}^2$ that minimize a metric \(\mathcal{L}(\boldsymbol{\theta})\) for the discrepancy between model predictions (parametric POD-based ROM or FOM) and some observational data capturing the diffusion-deformation process of the hydrogel.
To this end, material parameter identification can be cast as the below optimization problem:
\begin{equation}\label{eq:opt_problem_final}
    \boldsymbol{\theta}^{\text{opt}} = \text{arg}\min_{\boldsymbol{\theta}} \ \mathcal{L}(\boldsymbol{\theta}).
\end{equation}
The discrepancy metric \(\mathcal{L}(\boldsymbol{\theta})\) is often referred to as loss or objective function. In order to ensure proper convergence of the optimization problem stated above, the loss contributions from the different physical fields are computed and scaled individually. This leads to the following composite loss function
\begin{equation}\label{eq:loss_function}
    \mathcal{L} (\boldsymbol{\theta}) = \mathcal{L}_{\text{rel}}^{\mathbf{u}} (\boldsymbol{\theta}) + \mathcal{L}_{\text{rel}}^{\mu} (\boldsymbol{\theta}), 
\end{equation}
whose contributions are computed from the relative $L^2$ norms of the difference between observed and computed displacement fields $\bm{u}_{\text{obs}}$ and $\bm{u}(\bm{\theta})$
\begin{equation}
    \begin{aligned}
        \mathcal{L}_{\text{rel}}^{\mathbf{u}} &= \frac{\|\bm{u}_{\text{obs}} - \bm{u}(\boldsymbol{\theta}) \|_2}{\|\bm{u}_{\text{obs}} \|_2},
    \end{aligned}
\end{equation}
as well as of the difference between observed and computed concentration fields $\mu_{\text{obs}}$ and $\mu(\bm{\theta})$:
\begin{equation}
\mathcal{L}_{\text{rel}}^{\mu} = \frac{\| \mu_{\text{obs}} - \mu(\boldsymbol{\theta}) \|_2}{\| \mu_{\text{obs}} \|_2}.
\end{equation}

The solutions $\mu(\boldsymbol{\theta})$ and $\bm{u}(\boldsymbol{\theta})$ can be obtained either from the FOM predictions or its ROM approximation given by equation \eqref{eq:FOMbyROM} at those discrete time steps $t_i$ at which observational data is available. The solution of the optimization problem stated in equation \eqref{eq:opt_problem_final} requires iterative solution techniques, which makes the use of a ROM appealing. {In this work, the L-BFGS-B optimizer is used, a variant of the Limited Memory Broyden–Fletcher–Goldfarb–Shanno (L-BFGS) algorithm, which belongs to the family of quasi-Newton methods \citep{Zhu1997LBFGSB}. This algorithm is particularly suited for solving large-scale optimization problems with bound constraints. Its hyperparameters are provided in Section~\ref{subsec:num_ex_optimization}.}

\section{Numerical simulations}
\label{sec:simulation_results}

In this section, we conduct two extensive numerical studies to evaluate the effectiveness of parametric POD-based ROMs in capturing the main dynamics of the FOM describing the diffusion-deformation process of hydrogels.
The motivation behind the numerical computations is to capture the step-by-step process of manufacturing hydrogels using bioprinting as functional materials in biomedical applications. 
Typically, hydrogel fabrication is done in two stages.
The first stage involves characterizing the material to be manufactured, whereas, in the second stage, a bioprinter can be used to fabricate more intricate structures, possibly capturing complexities closer to those in living tissues.

\begin{figure}[!htb]
\center{\includegraphics[width=0.7\textwidth]{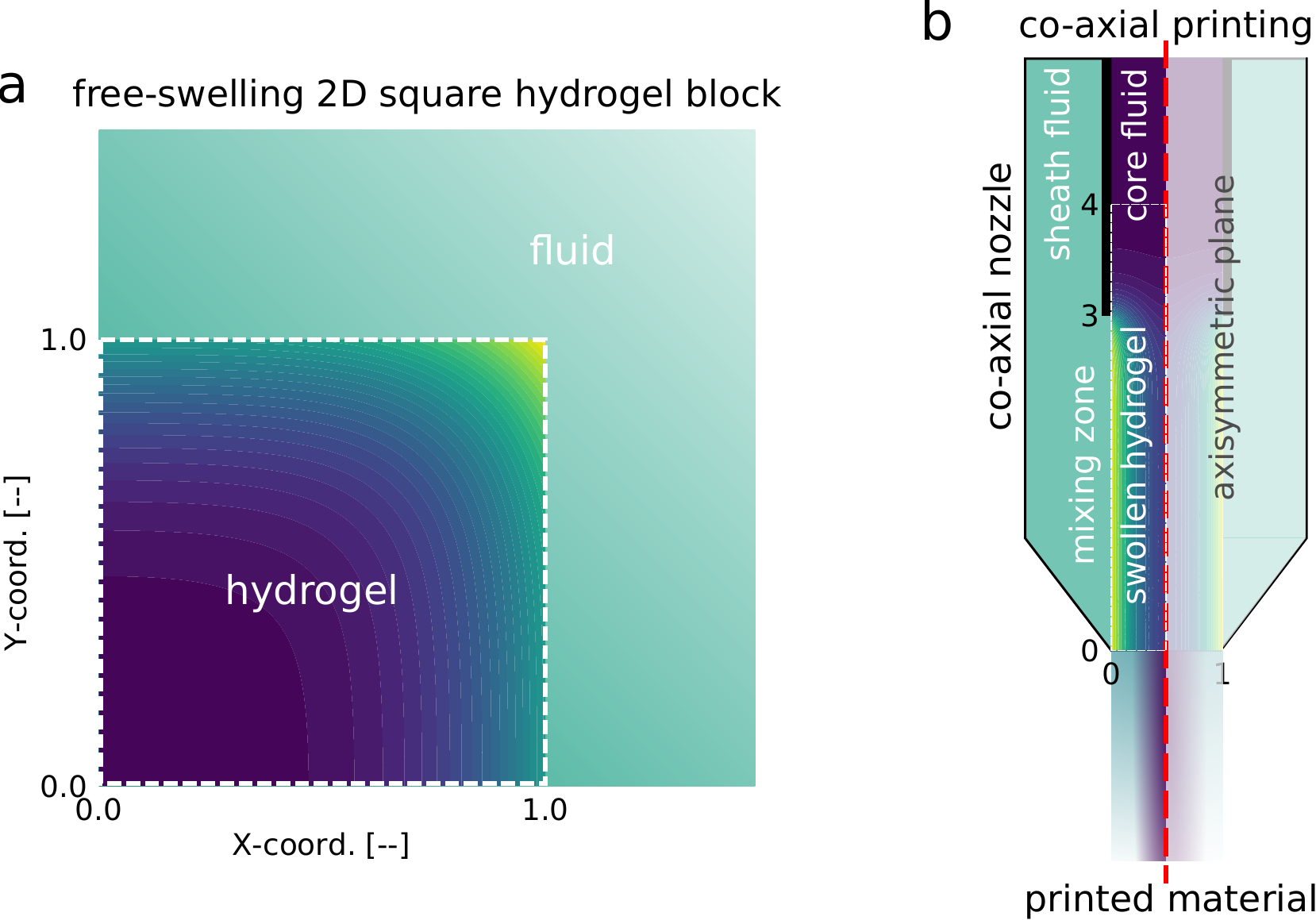}}
\caption{\textbf{Illustration of the simulated problems:} \textbf{a.} \textbf{benchmark:} 2D square hydrogel block and \textbf{b.} \textbf{case study:} co-axial printing. \textbf{Notes:} \textbf{i.} The white dashed boxes mark the domains extracted for the simulation setup. \textbf{ii.} The shadowed area in \textbf{b} refers to the neglected part due to the axisymmetric nature of the case study.}
\label{fig:simulation_setup}
\end{figure}

Figure \ref{fig:simulation_setup} illustrates the simulation setup of the two problems. First, a benchmark problem is considered, namely, a free-swelling of a 2D square hydrogel block (Figure \ref{fig:simulation_setup}\textbf{a}).
We implement parametric POD and nested-POD-based ROMs, test their accuracy, compare the simulation speed-up with respect to the FOM, and perform a material parameter identification to evaluate the ROMs' effectiveness in this many-query scenario.
The second test is a case study mimicking co-axial hydrogel bioprinting (Figure \ref{fig:simulation_setup}\textbf{b}). For the case study, we present the results for the parametric nested-POD-based ROM and test its accuracy and computation speed-up with respect to the FOM. The code for the POD-based ROM is provided for interested readers. Finally, we conduct a parameter uncertainty propagation and exploit the simulation speed-up achieved by the nested-POD-based ROM.
The FOM is solved using FEniCS\footnote{\url{https://fenicsproject.org/}} \citep{Alnaes2015Fenics} and the ROM is implemented in RBniCS\footnote{\url{https://www.rbnicsproject.org/}} \citep{Rozza2024RBniCsBook}. 
All codes to reproduce and verify the results can be accessed in the following online repository: \cite{Urrea2024MOR_hydrogels_code}.

\subsection{Benchmark: free-swelling of a 2D square hydrogel block}
\label{subsec:benchmark}

This subsection refers to the first stage of hydrogels manufacturing. 
The first stage involves characterizing the material to be manufactured. This is done through a series of experiments in the laboratory where molding techniques are used to fabricate simple hydrogel blocks under controlled conditions. These fabricated hydrogels are then subjected to, for example, free-swelling experiments to \textbf{i.} characterize material properties such as bulk and shear moduli and absorption capabilities of diffusing species and \textbf{ii.} validate mathematical models that can be used later as predictive tools for the material behavior. 

\subsubsection{Simulation setup}

As a benchmark problem, we consider the transient free-swelling of a gel in a two-dimensional setting as illustrated in Figure \ref{fig:simulation_setup}\textbf{a}. The 2D hydrogel block initially has a square cross-section.
The hydrogel block is immersed into a non-reactive solvent with a reference chemical potential $\mu^0 = 0$. Only a quarter of the whole model is considered because of the symmetry of the block. In terms of mechanical boundary conditions, the bottom and left edges are subject to Dirichlet boundary conditions $u_y = 0$ and $u_x = 0$, respectively. The top and right edges are considered to be traction-free. In terms of chemical boundary conditions, the bottom and left edges are subject to zero fluid flux. The chemical potential is subject to a Robin boundary condition as defined in equation \eqref{eq:robin_BC_final} on the top and right edges.
A similar setup to this example can be found in, e,g., \cite{Chester2011thermo}, \cite{Liu2016TransGels}, or \cite{Chen2020LinearHydrogelsExp}. 

The following values are selected as the nominal dimensionless material parameter\footnote{\textbf{Note:} The dimensionless material parameters $(\lambda, A)$ are related to the following physical material parameters through equations~\eqref{eq:normalized_parm}: $l = 0.0015$ [m], $k_B T = 4e-21$ [J], $\nu = 3e-29$ [m$^3$], $D^* = 1.6e-11$ [m$^2$/s], $G = 33$ [kPa], and $\lambda^{d} = 1600*G$ [kPa].}: $\lambda = 1558$ and $A = 4000$, and the Robin boundary coefficient is defined as $\alpha = 0.66$. The chemical potential's initial condition is $\mu_{o} = -0.3124$, and the domain is considered to be initially stress-free. These material parameters taken from \cite{Chen2020LinearHydrogelsExp} represent a gelatin-based hydrogel. 
The reader is referred to \cite{Chen2020LinearHydrogelsExp} for more details on the gel preparation and its mechanical characterization. 

The domain is discretized into triangular elements with quadratic interpolation for displacement and linear interpolation for the chemical potential to obtain the Taylor-Hood elements as introduced in Section \ref{sec:gels_theory}.
The dimensionless coupled system of equations \eqref{eq:disc_weak_form} is solved as a variational monolithic problem using the FEniCS package as a PDE solver. 
We remark here that we solve the FOM in a monolithic fashion, but later for the POD, we partition the snapshot matrix into two parts in order to apply the partitioned approach as described in Section \ref{sec:MOR}.

Following the convergence analysis reported in \ref{app:conv_analysis}, a mesh density of $N_h = 50$, that is, $5000$ triangular elements, is specified to obtain what we denote as the FOM nominal simulation setup. The total dimensionless simulation time is chosen as $T = 0.25$, and the time interval $\lbrace 0,~T \rbrace$ is discretized into $100$ time steps. The reader should notice that $T = 0.25$ corresponds to about $t = 3$~hours of the hydrogel diffusion-deformation before normalization based on equation~\eqref{eq:normalized_parm}. 
In their experimental work, \cite{Chen2020LinearHydrogelsExp} found that the linear model is valid up to $3$~h of fluid absorption. For larger simulation times, a nonlinear theory should be adopted \citep{Urrea2023DiffDef_hydrogels}.

\subsubsection{Numerical results for the full-order model}

\begin{figure}[!htb]
\center{\includegraphics[width=0.95\textwidth]{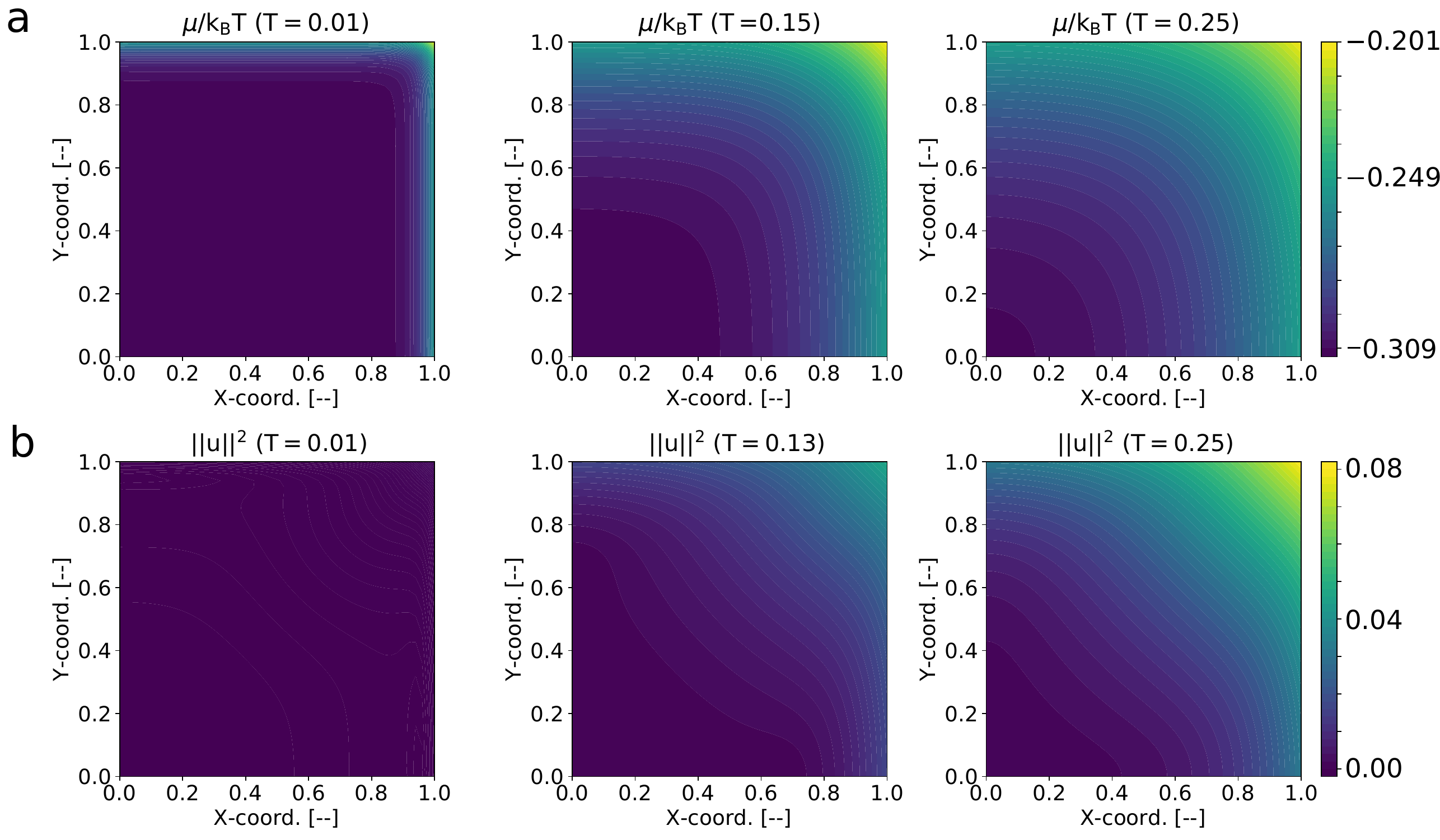}}
\caption{\textbf{Benchmark: FOM simulation results for primary variables.} \textbf{a.} normalized chemical potential and \textbf{b.} displacement magnitude at time steps $T = \{0.01, 0.15, 0.25\}$.}
\label{fig:FOM_mu_disp}
\end{figure}

Figure \ref{fig:FOM_mu_disp} shows the primary variables for the FOM at three different simulation times. Figure \ref{fig:FOM_mu_disp}\textbf{a} displays the evolution of the chemical potential within the 2D hydrogel block. It can be observed that the chemical potential transitions over time from its initial condition to its boundary value. This increase in chemical potential indicates that as the gel absorbs the solvent, its internal fluid concentration rises, driving the absorption process towards equilibrium.

Figure \ref{fig:FOM_mu_disp}\textbf{b} illustrates the evolution of the displacement magnitude. It is evident that displacement increases in tandem with solvent absorption, with the highest displacement occurring in the corner fully exposed to the solvent. 
The symmetry in the displacement field reflects the isotropic swelling of the gel.

\begin{figure}[!htb]
\center{\includegraphics[width=0.95\textwidth]{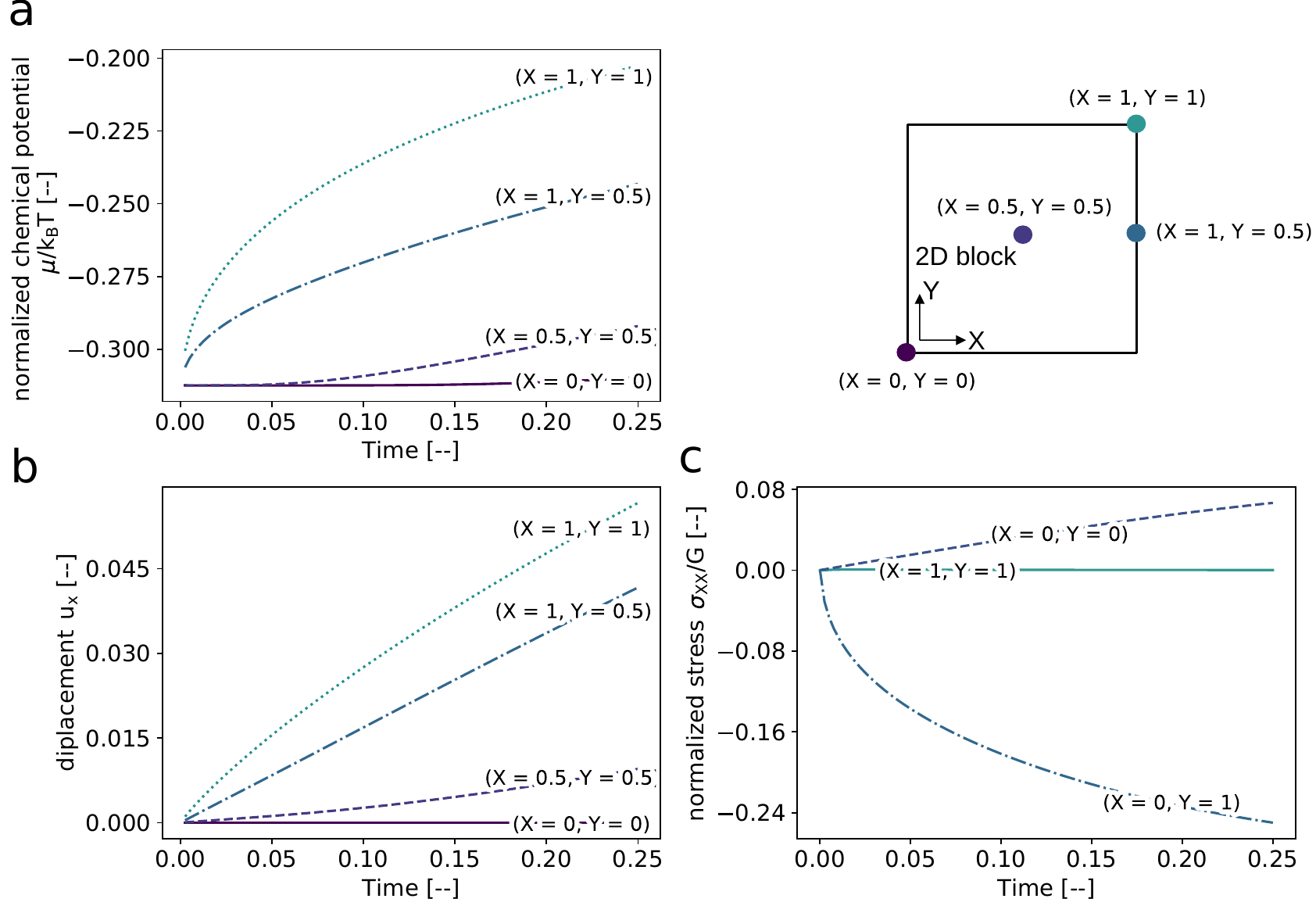}}
\caption{\textbf{Benchmark: transient FOM simulation.} \textbf{a.} normalized chemical potential, \textbf{b} displacement in x-direction, and \textbf{b.} normalized stresses at different points in the 2D hydrogel block.}
\label{fig:FOM_time}
\end{figure}

Figure \ref{fig:FOM_time} shows the time evolution of normalized chemical potential (Figure \ref{fig:FOM_time}\textbf{a}), displacement (Figure \ref{fig:FOM_time}\textbf{b}), and normalized stress (Figure \ref{fig:FOM_time}\textbf{c}) at different points in the 2D hydrogel block. The figure reveals that the four selected points undergo different diffusion-deformation pathways, highlighting the spatial dependency of the problem. Figure \ref{fig:FOM_time} complements observations from Figure \ref{fig:FOM_mu_disp}. 
Figure \ref{fig:FOM_time}\textbf{c} shows that while a region within the domain experiences compressive stresses, another region experiences the opposite. This stress gradient is the main driver of the fluid's diffusion. The figure also reaffirms that the points on the right side remain traction-free throughout the simulation.

\subsubsection{Numerical results for the reduced-order model}

In this section, we provide numerical experiments to validate and demonstrate the ROMs' accuracy and computational efficiency. All the following computations are conducted in RBniCS. The same mesh density and time steps as for the nominal simulation are taken. Material parameters are sampled within the ranges 
$\lambda=\left[1000, 2000 \right]$ and $A = \left[2000, 6000\right]$.

In order to minimize the overall computational effort, the immediate task is now to identify the minimum number of basis functions and samples. For this purpose, we follow the procedure presented in Section~\ref{sec:MOR} and summarized in Figure~\ref{fig:POD_workflow_illustration}.
This procedure is followed for both the traditional parametric POD-based ROM and the nested variant, both of which have been introduced in Section~\ref{sec:MOR}. 

\begin{figure}[!htb]
\center{\includegraphics[width=0.95\textwidth]{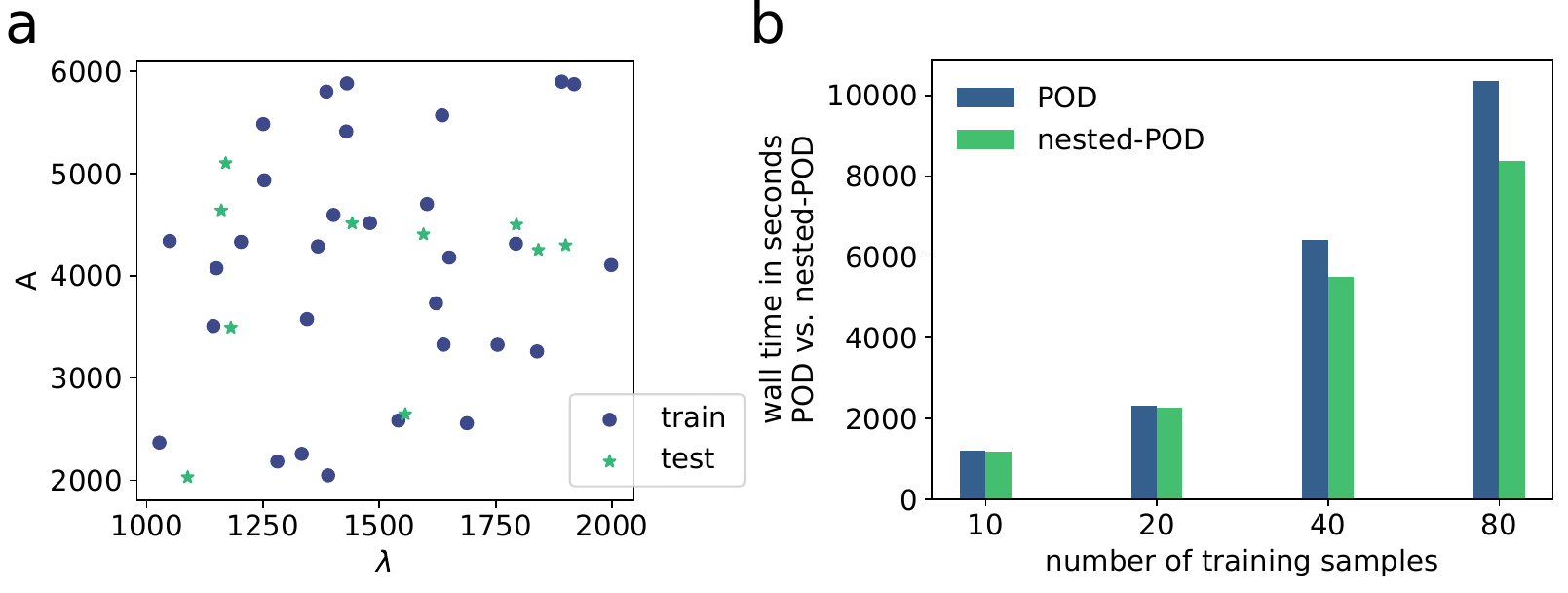}}
\caption{\textbf{Benchmark:} material parameter sampling for the POD and nested-POD-based ROM construction.}
\label{fig:2D_square_ROM_param_sampling_wall_time}
\end{figure}

Figure \ref{fig:2D_square_ROM_param_sampling_wall_time}\textbf{a} depicts the generated training and testing samples from the material parameters used in the construction of the ROM. The samples were generated using Monte Carlo sampling from a uniform distribution.
We noticed that 10 samples are the minimum to obtain an acceptable approximation of the FOM with both POD and nested-POD-based ROM. We trained the POD-based ROMs using 30 samples.
Figure~\ref{fig:2D_square_ROM_param_sampling_wall_time}\textbf{b} reports the wall time required for training the ROMs adopting the POD and nested-POD algorithms for an increasing number of training samples. The wall time is composed of the time required to create the snapshots and to perform the SVD on the resultant snapshots matrix.

\begin{figure}[!htb]
\center{\includegraphics[width=1.0\textwidth]{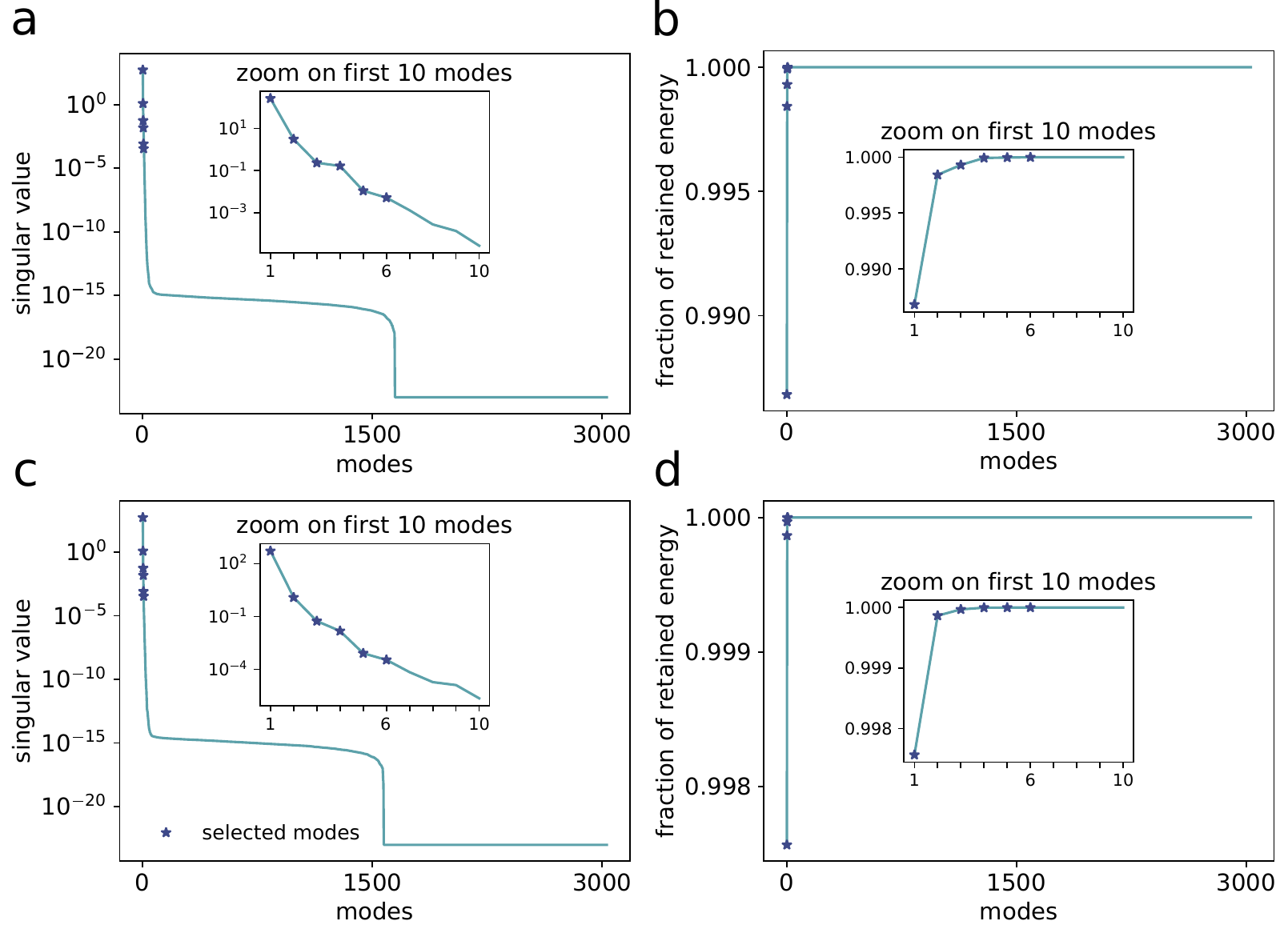}}
\caption{\textbf{Benchmark: POD-based ROM.} SVD of the snapshots matrices for \textbf{chemical potential:} \textbf{a.} singular values and \textbf{b.} retained energy; and \textbf{displacement:} \textbf{c.} singular values and \textbf{d.} retained energy.}
\label{fig:2D_square_POD_truncation}
\end{figure}

\textbf{POD-based approach:} Figure~\ref{fig:2D_square_POD_truncation} shows the singular values and energy retained by the SVD on the snapshots matrix.
From Figures~\ref{fig:2D_square_POD_truncation}\textbf{a} and \textbf{c}, it is observed that the singular values magnitude decays very fast for both chemical potential and displacement. Such a behavior is advantageous, as it facilitates the development of an efficient low-dimensional ROM. In fact, from Figures~\ref{fig:2D_square_POD_truncation}\textbf{b} and \textbf{d}, it is possible to conclude that only six basis functions are required to retain up to $\eta = 99.9999 \%$ of the energy contained in the FOM. 

\begin{figure}[!htb]
\center{\includegraphics[width=1.0\textwidth]{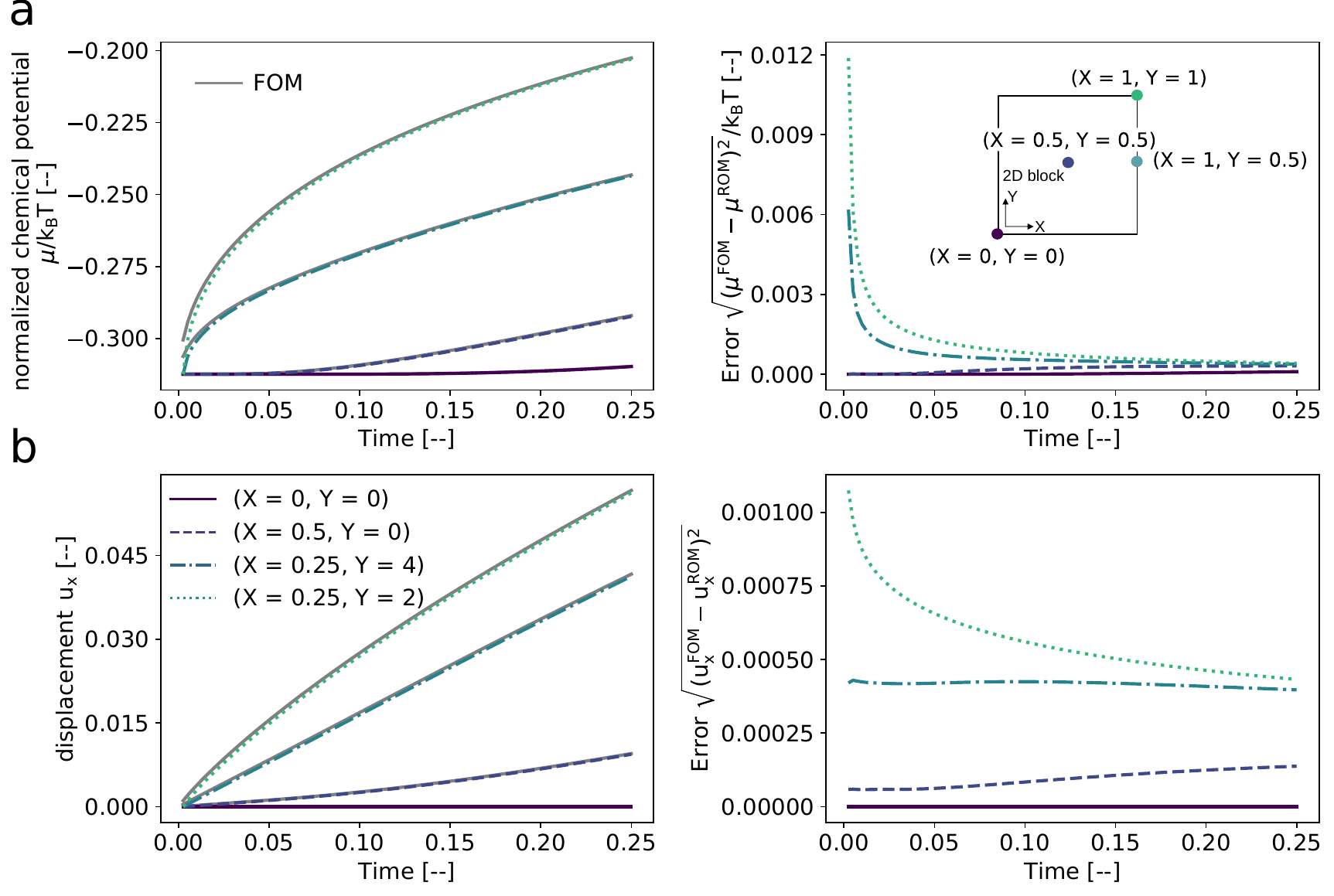}}
\caption{\textbf{Benchmark: POD-based ROM simulation results with six basis functions.} \textbf{a.} normalized chemical potential and square root error with respect to the FOM; \textbf{b.} displacement magnitude at different simulation times, and square root error with respect to the FOM.}
\label{fig:POD_mu_disp_errors}
\end{figure}

An illustration of the error for the selected number of basis functions can be found in Figure~\ref{fig:POD_mu_disp_errors}. Here, the POD is assembled, and the predictions are compared to the nominal FOM model ($A=4000, \lambda=1558$, and $\alpha=0.66$). The reader should notice that these parameter values were not included in the training dataset; therefore, they can be used to validate the POD-based ROM. 

Figure~\ref{fig:POD_mu_disp_errors} depicts a comparison of the temporal evolution of chemical potential (Figure~\ref{fig:POD_mu_disp_errors}\textbf{a}) and displacement (Figure~\ref{fig:POD_mu_disp_errors}\textbf{b}) at different points inside the 2D hydrogel block for the POD-based ROM. The figures show that the root mean square error (RMSE) between the POD-based ROM and FOM has its maximum at the beginning of the simulation.
However, for the chemical potential, the RMSE reaches $3 \times 10^{-3}$ in about $T = 0.05$ and keeps decreasing as time evolves. For the displacement, the RMSE is smaller than $10^{-3}$ for almost the whole simulation time.  

\begin{figure}[!htb]
\center{\includegraphics[width=1.0\textwidth]{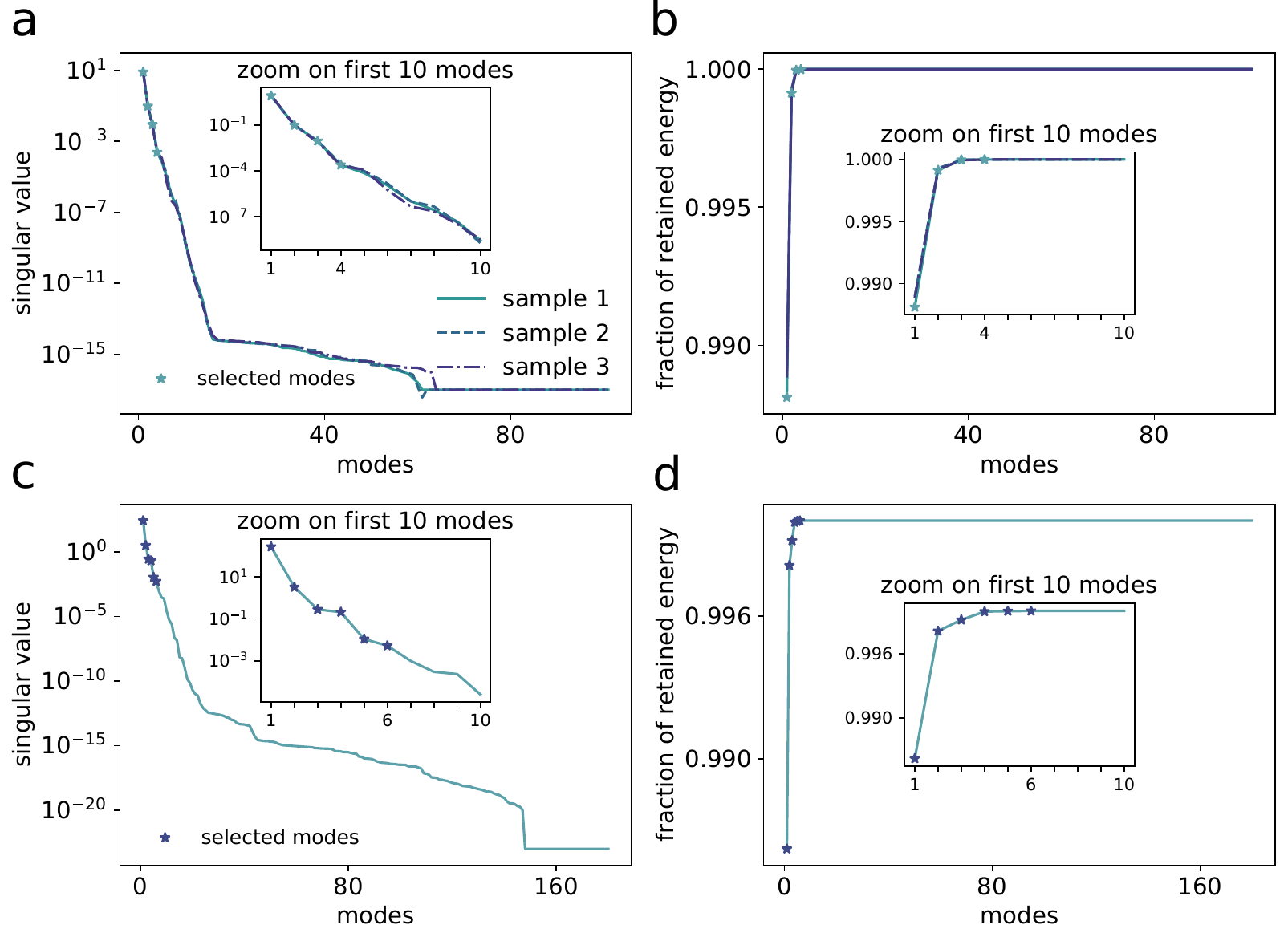}}
\caption{\textbf{Benchmark: nested-POD-based ROM.} SVD of the \textbf{chemical potential} snapshots matrix: \textbf{a.} singular values and \textbf{b.} retained energy for the temporal domain; and \textbf{c.} singular values and \textbf{d.} retained energy for the spatial domain.}
\label{fig:2D_square_nestedPOD_truncation_mu}
\end{figure}

\begin{figure}[!htb]
\center{\includegraphics[width=1.0\textwidth]{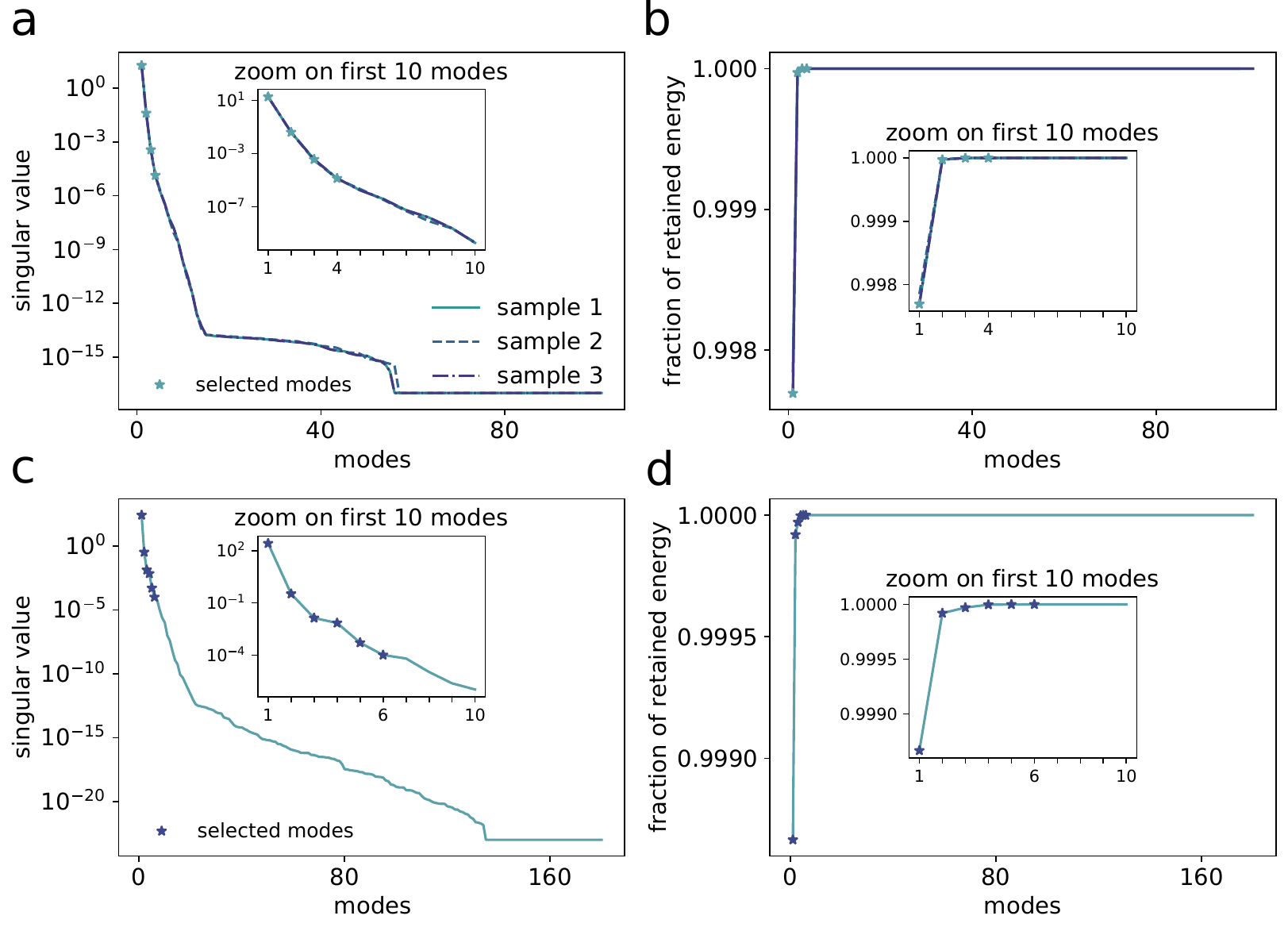}}
\caption{\textbf{Benchmark: nested-POD-based ROM.} SVD of the \textbf{displacement} snapshots matrix: \textbf{a.} singular values and \textbf{b.} retained energy for the temporal domain; and \textbf{c.} singular values and \textbf{d.} retained energy for the spatial domain.}
\label{fig:2D_square_nestedPOD_truncation_disp}
\end{figure}

\textbf{Nested-POD-based approach:} Figures~\ref{fig:2D_square_nestedPOD_truncation_mu} and \ref{fig:2D_square_nestedPOD_truncation_disp} report the singular values and fraction of energy retained for the chemical potential and displacement for the nested-POD approach, respectively. On the one hand, Figures~~\ref{fig:2D_square_nestedPOD_truncation_mu}\textbf{a}, \textbf{b} and \ref{fig:2D_square_nestedPOD_truncation_disp}\textbf{a}, \textbf{b} report the behavior of the singular values and retained energy for the temporal dimension, for which SVD is locally applied for each material parameter sample. The behavior of the singular values and retained energy are presented for three out of the 30 samples taken for the parameters. The figures show that the singular values and retained energy behavior are very similar for the temporal dimension in each case. From the analysis, it is concluded that only four temporal modes are necessary to retain the desired fraction of energy $\eta$.
On the other hand, Figures~\ref{fig:2D_square_nestedPOD_truncation_mu}\textbf{c}, \textbf{d} and \ref{fig:2D_square_nestedPOD_truncation_disp}\textbf{c}, \textbf{d} display the behavior of the singular values and retained energy for the spatial dimension, for which SVD is applied over the snapshots matrix containing the temporal basis functions and spatial information for the 30 training parameter samples. From this analysis, it is concluded that only six spatial basis functions are required to retain the desired fraction of energy $\eta$ contained in the FOM for the chemical potential and displacement.

\begin{figure}[!ht]
\center{\includegraphics[width=0.95\textwidth]{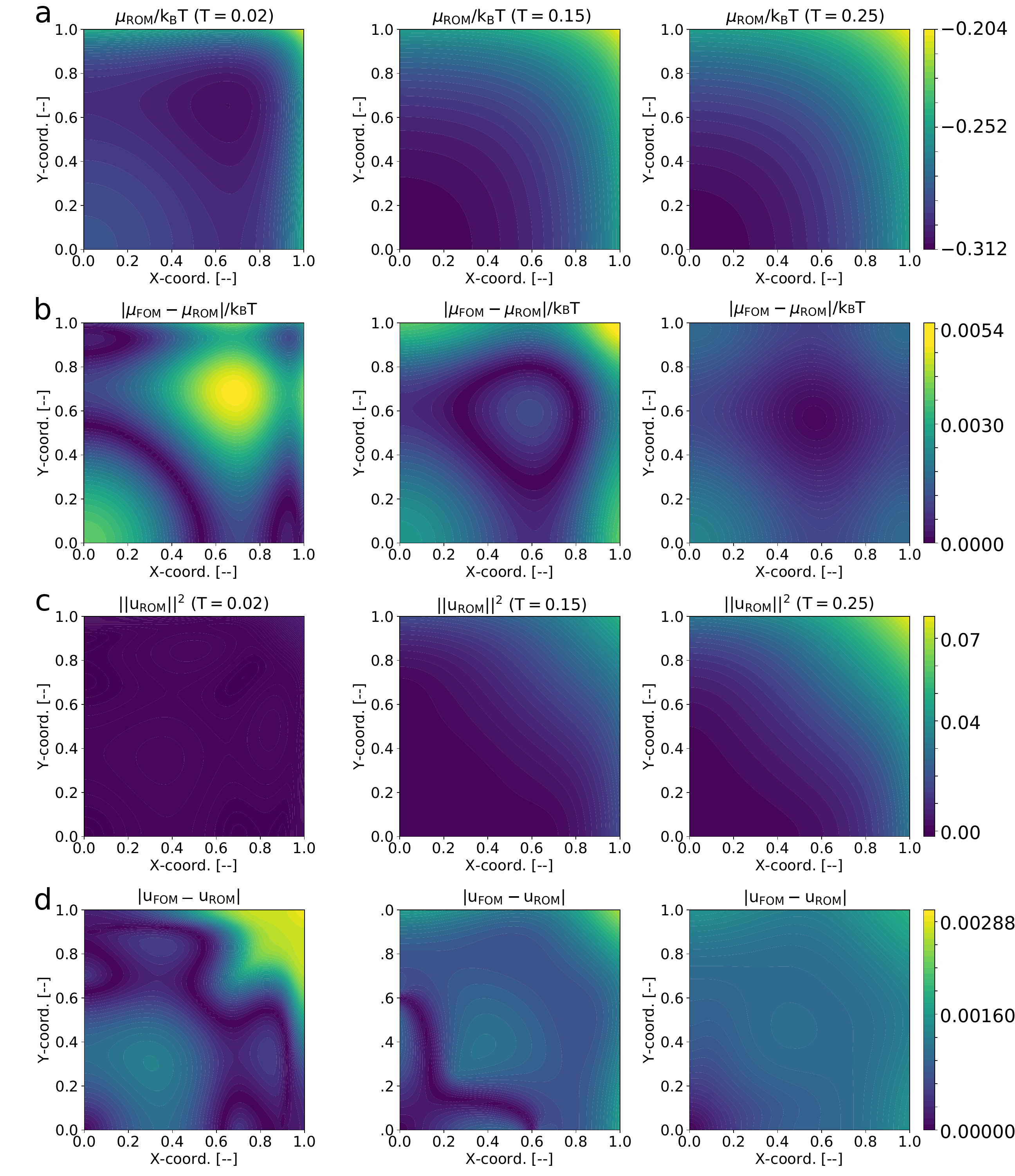}}
\caption{\textbf{Benchmark: nested-POD-based ROM simulation results for primary variables.} \textbf{a.} normalized chemical potential field and \textbf{b.} absolute ROM error with respect to the FOM, as well as \textbf{c.} displacement magnitude field and \textbf{d.} absolute ROM error with respect to the FOM at time steps $T = \{0.01, 0.15, 0.25\}$.}
\label{fig:nestedPOD_ROM_mu_disp}
\end{figure}

As we found that the results obtained with the nested-POD-based ROM were identical to those reported in Figure~\ref{fig:POD_mu_disp_errors}, we instead plot a full-field visualization of the simulation outcome for $[\lambda, A] = [1558, 4000]$.
To this end, Figure~\ref{fig:nestedPOD_ROM_mu_disp} displays the chemical potential (Figure~\ref{fig:nestedPOD_ROM_mu_disp}\textbf{a}) and displacement (Figure~\ref{fig:nestedPOD_ROM_mu_disp}\textbf{c}) obtained with the six basis functions at three different time steps for the nested-POD-based ROM, namely, $T= \{0.02, 0.15, 0.25\}$.
Errors with respect to the FOM simulation results reported in Figure~\ref{fig:FOM_mu_disp} are displayed for the chemical potential in Figure~\ref{fig:nestedPOD_ROM_mu_disp}\textbf{b} and for the displacement in Figure~\ref{fig:nestedPOD_ROM_mu_disp}\textbf{d}. 
As seen from Figure~\ref{fig:nestedPOD_ROM_mu_disp}, the nested-POD-based ROM can predict the FOM to a satisfactory extent. From Figures~\ref{fig:nestedPOD_ROM_mu_disp}\textbf{b} and \textbf{d}, it can be observed that the maximum discrepancy between the FOM and ROM is less than $0.55\%$ for the chemical potential and $0.3\%$ for the displacement field, respectively. It is also clear that the maximum discrepancy occurs at the beginning of the simulation but fades as time evolves. 
This result is expected because, at the beginning of the simulation, the chemical potential field has a very sharp gradient close to the exposed boundaries, but it is almost flat towards the center. It is well known that global surrogate models suffer when approximating flat functions.

POD and nested POD-based ROMs require the same number of basis functions to retain the desired fraction of energy $\eta$, and both approaches lead to the same level of accuracy; see also \ref{app:conv_analysis_ROM} for the convergence error analysis. However, the demand for computational resources significantly increases with the number of parameter samples. The advantage of the nested-POD algorithm becomes evident as reported in Figure~\ref{fig:2D_square_ROM_param_sampling_wall_time}\textbf{b}. While both algorithms perform similarly with a small number of training samples, the nested-POD algorithm becomes more efficient as the number of training samples increases.

\begin{figure}[!htb]
\center{\includegraphics[width=0.45\textwidth]{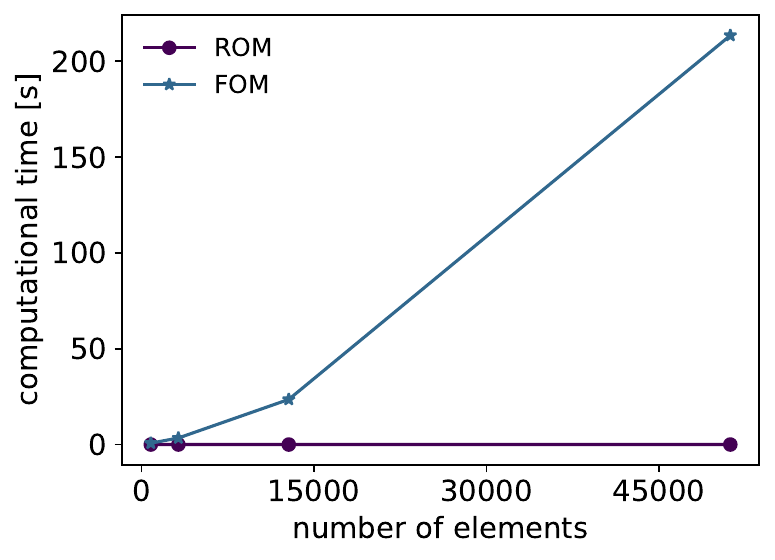}}
\caption{\textbf{Benchmark: nested-POD-based ROM computational speed-up compared to the FOM.}}
\label{fig:nestedPOD_ROM_cpu_speedup}
\end{figure}

Figure \ref{fig:nestedPOD_ROM_cpu_speedup} demonstrates the significant computational speed-up achieved with the ROM. 
For instance, the FOM computation time for $N_h = 10$ is $0.33$s, but for $N_h = 160$, it is already $213.4$~s. This corresponds to an exponential increase of the computational cost w.r.t. the mesh density. In contrast, the ROM computation time for $N_h = 10$ is $0.018$~s, but for $N_h = 160$, it is still only $0.025$~s. That is, the ROM simulation is about a factor of $8536$ faster than the FOM simulation for $N_h = 160$.

In the following, we restrict our analysis to the nested-POD approach only. This is because the same number of basis functions is required to retain the desired fraction of energy $\eta$, the same level of accuracy has been obtained with POD and nested POD-based ROMs, and given the computational advantage of nested-POD algorithm for a large number of training samples. However, the code is provided to replicate results with POD and nested-POD-based ROM for all upcoming examples; see \cite{Urrea2024MOR_hydrogels_code}.

\subsubsection{Material parameter identification from full-field data}
\label{subsec:num_ex_optimization}

{This section shows the potential of using the developed nested-POD-based ROM to identify the model's material parameters from full-field data. In the lack of experimental data, we take the solution of the nominal FOM ($\lambda=1558, A=4000$) at $T = \{0.15, 0.25\}$ as synthetic data.
It is worth recalling that the ROM did not encounter these values in the training dataset. 

The optimization problem \eqref{eq:opt_problem_final} to identify the optimal value of the material parameters was implemented in Python and solved using the minimize function from the Scipy library \citep{Virtanen2020scipy}. The hyperparameters of the L-BFGS-B algorithm are the maximum number of function evaluations and iterations (`maxfun' and `maxiter'), both set to $15,000$. The function tolerance (`ftol') is set to $2.22 \times 10^{-9}$, and the gradient tolerance (`gtol') to $1 \times 10^{-5}$. The function tolerance dictates the precision of the objective function and the convergence criterion based on the gradient's magnitude, respectively. The step size for numerical approximation (`eps') is $1 \times 10^{-8}$, and the maximum number of line search steps (`maxls') per iteration is $20$. This setup allowed for efficient parameter space exploration while adhering to the constraints imposed by the problem's physical context.

The optimization process starts with an initial guess for the material parameters, set to $\bm{\theta}_0 = [\lambda_0, A_0] = [1800, 3800]$, but we verified that the same result is obtained if different initial conditions are chosen. The bound constraints $[\lambda, A] = [(1000, 2000), (2000, 6000)]$  ensure that the solution remains within the parameter range the ROM was trained for.}

Table \ref{tab:param_identification_ROMvsFOM} summarizes the results obtained from the parameter identification with ROM and FOM, as well as the related computational cost. 
From the results, it is possible to observe the nested-POD-based ROM's effectiveness in performing many-query tasks. The material parameters identified with the ROM have a relative error of $0.95~\%$ and $1.95~\%$ for $\mu$ and $\mathbf{u}$, respectively. This discrepancy is attributed to the low number of temporal snapshots $(T = \{0.15, 0.25\})$ used for the identification problem. If more snapshots were considered for the identification problem, the error between the identified and correct material parameters is expected to decrease, but especially with the FOM, the computational cost would also increase significantly. Considering both the ROM training and the actual parameter identification, only a third of the time needed with the FOM was consumed. It is important to remark that the material parameter values identified using the FOM also deviate from the nominal values used to create the synthetic data. 

\begin{table}[]
    \centering
    \begin{tabular}{r|l|l}
                                                                                & \textbf{ROM}               & \textbf{FOM} \\
                                            \hline
       ROM model eval. only training (computation time)                    & 0 (0h)                     & 30 (30min)      \\
       model eval. only identification (computation time)          & 90 (2.5min)                 & 84 (90min)     \\
       computation time training + identification                                & 30min + 5min                  & n.a. + 90min      \\ 
       \hline
       \textbf{dimensionless first Lamé constant} $\lambda$ & 1572 & 1542 \\
       absol. error $\vert \lambda - \lambda^{\text{opt}} \vert$ & 14 & 30 \\
       relative error [\%] & 0.90 & 1.95 \\
       \hline
       \textbf{dimensionless scaling factor $A$} & 3872 & 3967 \\
       absol. error $\vert A - A^{\text{opt}} \vert$ & 128 & 95 \\
       relative error [\%] & 3.20 & 2.39 \\
       \hline
    \end{tabular}
    \caption{\textbf{Benchmark: nested-POD-based ROM v.s. the FOM.} Computational analysis of the model parameters identification.}
    \label{tab:param_identification_ROMvsFOM}
\end{table}

It is worth remarking that the optimization in equation \eqref{eq:opt_problem_final} converges within 6 steps using the FOM, but 84 model calls were necessary to approximate the Jacobian. Thus, the computational time could be reduced significantly when using adjoints. However, our focus is on repeated calibration in laboratory contexts, where the ROM would still be beneficial.

\subsection{Case study: co-axial hydrogel bioprinting}

Co-axial bioprinting is one technique that can be used to automate the fabrication process \citep{Kjar2021coaxial_print}. 
It uses a dual-nozzle in a core/sheath configuration system to extrude a hydrogel precursor and a crosslinking agent simultaneously; see Figure \ref{fig:simulation_setup}\textbf{b}. This core/sheath configuration, which enables immediate gelation and structural stability, can be adapted for specific hydrogel composition designs, allowing for multi-material deposition and improved resolution. It is crucial in tissue engineering for creating complex 3D structures with controlled properties like stiffness and porosity \citep{Kjar2021coaxial_print}.

Model validation in the first stage is then employed to make predictions of the co-axial printing process and understand the effect of material parameter uncertainties on the printed hydrogel.
In this case, the model might need to be evaluated many times to evaluate different printing conditions. 
More precisely, the material parameters $\lambda$ and $A$ could present a significant source of uncertainty. 
Consequently, in Subsection \ref{subsec:unc_prop}, we exploit the ROM's efficiency and propagate the uncertainty in the material parameters onto two quantities of interest (QoIs) in the co-axial process, namely, the chemical potential at the nozzle's tip and the induced stresses within the domain due to swelling. 

On the one hand, the chemical potential is directly related to the species concentration and can inform about the degree of crosslinking (DoC) of the extruded material \citep{Hajikhani2021chemomechanics}. The DoC is a QoI that influences the gel's properties and determines its possible application.
On the other hand, induced stresses due to fluid absorption at the co-axial printer's nozzle are often overlooked in the literature on hydrogel bioprinting. The focus has primarily been on induced shear stresses from extrusion \citep{Chirianni2024BioprintingStress}. However, this does not capture the full picture of co-axial printing, where additional stresses are induced in the nozzle due to fluid absorption and hydrogel crosslinking, resulting in swelling and shrinking stresses. 
The mean and maximum induced stresses could be monitored to evaluate if the printed hydrogel has the right conditions for cells viability after printing \citep{Chirianni2024BioprintingStress}.

\subsubsection{Simulation setup}

We consider a vertical hydrogel bar where the height is four times the length as marked by the dashed white box in Figure \ref{fig:simulation_setup}\textbf{b}. 
We assume that the diffusion time of the fluid into the hydrogel is relatively small compared to the velocity at which the hydrogel is extruded out of the printer's nozzle. This allows us to neglect the physics of the hydrogel transport through the co-axial nozzle.
We study the transient free-swelling of the hydrogel bar in a two-dimensional setting, assuming a homogeneous mix of gel and solvent at the nozzle’s tip.
The material parameters are the same as in the benchmark problem mentioned earlier. Therefore, if the characteristic length in the previous example is $l = 0.0015$~m, the dimensions of the hydrogel bar are $0.0015 \times 0.0060$~m$^2$ or, equivalently, $1.5 \times 6.0$~mm$^2$. This size is representative of a filament being crosslinked in the nozzle of a co-axial bioprinter.

We assume that only $75 \%$ of the gel bar is exposed to the solvent on the left and right edges.
The remaining $25 \%$ of the left and right edges and the top surface have zero fluid flux for the solvent concentration boundary conditions. 
Due to the symmetry of the problem, only half of the domain is considered. The axisymmetric axis is marked in red in Figure~\ref{fig:simulation_setup}\textbf{b}. 
A Robin boundary condition imposes the chemical potential on the exposed portion on the left edge as defined in equation \eqref{eq:robin_BC_final}. 
Symmetric (Neumann) boundary conditions are defined at the center of the domain.
In terms of mechanical boundary conditions, the axisymmetric edge is subject to a Dirichlet boundary condition $u_x = 0$, while all other boundaries are considered to be traction-free.

The domain is discretized into triangular elements with quadratic interpolation for displacement and linear interpolation for the chemical potential, resulting in Taylor-Hood elements as introduced in Section \ref{sec:gels_theory}. Equation \eqref{eq:disc_weak_form} is solved as a variational monolithic problem using the FEniCS package as a PDE solver.

\subsubsection{Numerical results for the reduced-order model}

In this section, we provide numerical experiments to validate and demonstrate the parametric nested-POD-based ROM's accuracy and computational efficiency, as we did for the benchmark case. The procedure in Figure~\ref{fig:POD_workflow_illustration} is also followed in this section to assemble the nested-POD-based ROM.
The material parameters are varied within the same range as in the benchmark problem to create the training and testing sets; see Figure \ref{fig:2D_square_ROM_param_sampling_wall_time}.

\begin{figure}[!htb]
\center{\includegraphics[width=1.0\textwidth]{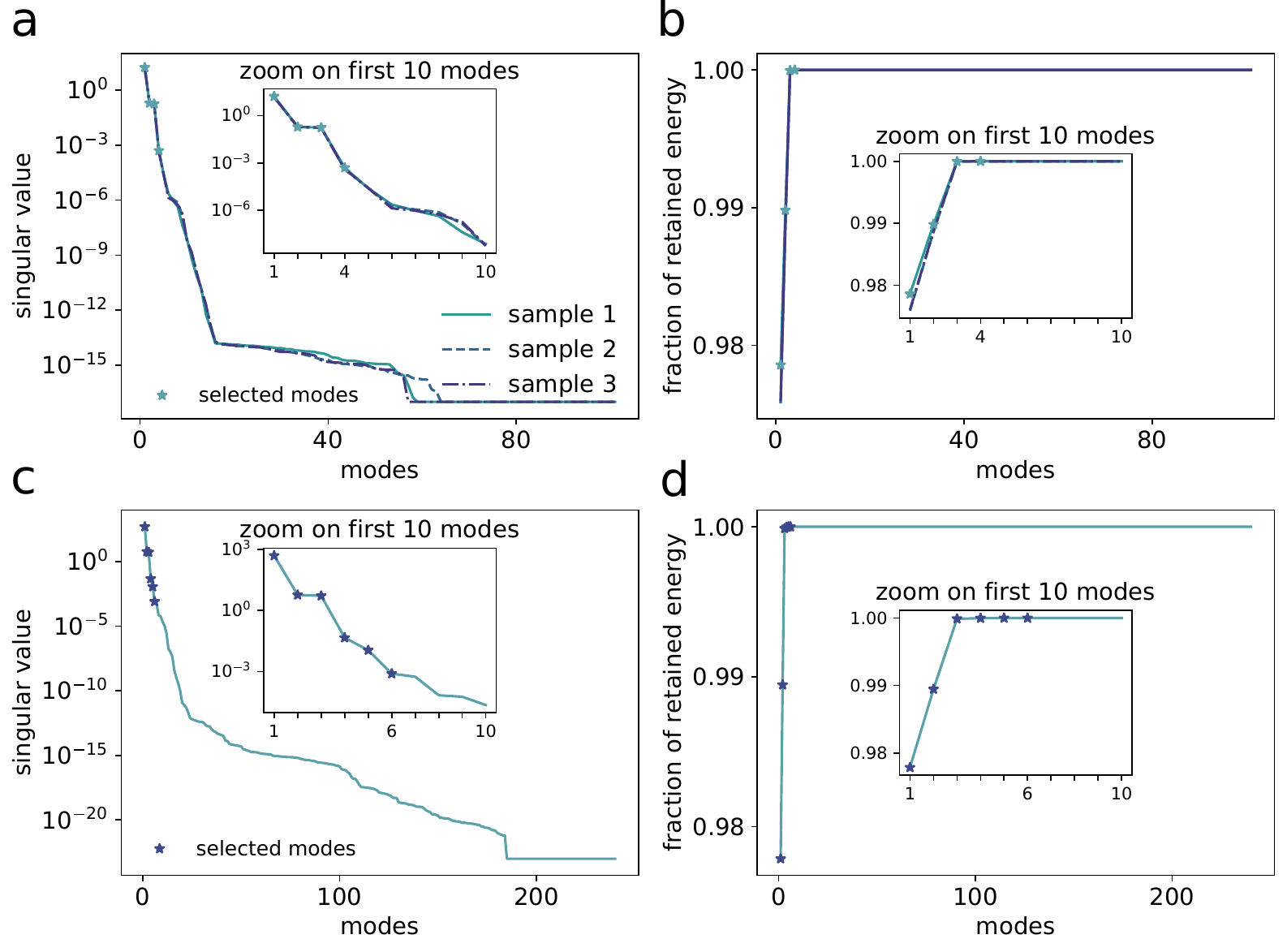}}
\caption{\textbf{Case study: hydrogel bar - nested-POD-based ROM.} SVD of the \textbf{chemical potential} snapshots matrix: \textbf{a.} singular values and \textbf{b.} retained energy for the temporal domain; \textbf{c.} singular values and \textbf{d.} retained energy for the spatial domain.}
\label{fig:gel_bar_nestedPOD_truncation_mu}
\end{figure}

\begin{figure}[!htb]
\center{\includegraphics[width=1.0\textwidth]{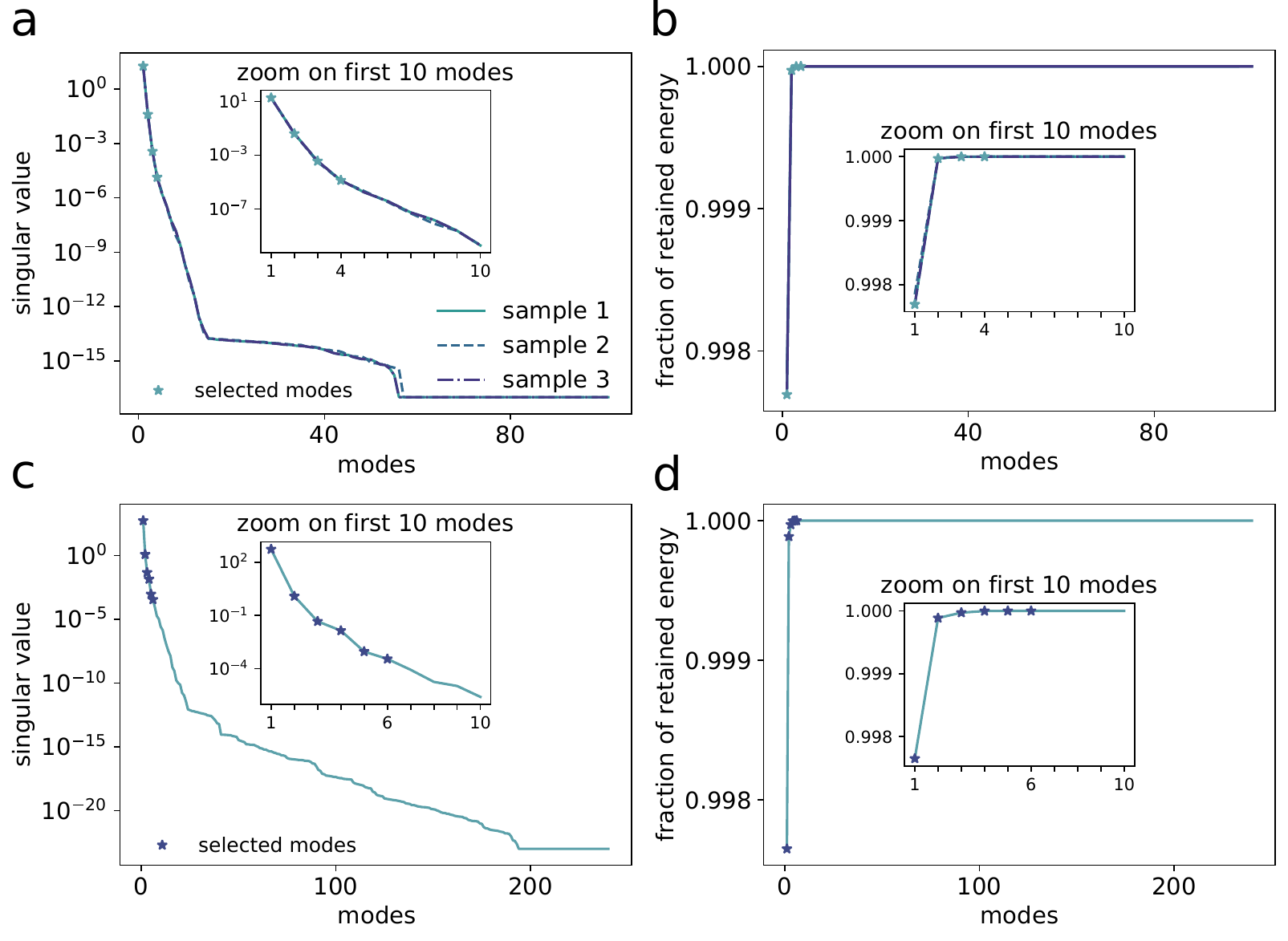}}
\caption{\textbf{Case study: hydrogel bar - nested-POD-based ROM.} SVD of the \textbf{displacement} snapshots matrix: \textbf{a.} singular values and \textbf{b.} retained energy for the temporal domain; \textbf{c.} singular values and \textbf{d.} retained energy for the spatial domain.}
\label{fig:gel_bar_nestedPOD_truncation_disp}
\end{figure}

Figures~\ref{fig:gel_bar_nestedPOD_truncation_mu} and \ref{fig:gel_bar_nestedPOD_truncation_disp} show the singular values and fraction of energy retained for the nested-POD algorithm for the chemical potential and displacement, respectively. On the one hand, Figures~\ref{fig:gel_bar_nestedPOD_truncation_mu}\textbf{a}, \textbf{b} and \ref{fig:gel_bar_nestedPOD_truncation_disp}\textbf{a}, \textbf{b} report the behavior of the singular values and retained energy for the temporal dimension. They show the decay of the singular values and the retained energy for three out of the 30 parameter samples, as we did for the benchmark problem. It is seen from the figures that the singular values and retained energy behavior are very similar for the temporal dimension. We conclude from the analysis that six temporal modes are enough to retain $\eta=99.9999 \%$ of the energy.
On the other hand, Figures~\ref{fig:gel_bar_nestedPOD_truncation_mu}\textbf{c}, \textbf{d} and \ref{fig:gel_bar_nestedPOD_truncation_disp}\textbf{c}, \textbf{d} display the decay of the singular values and retained energy for the spatial dimension. This analysis concludes that we can retain the desired fraction of energy $\eta$ contained in the FOM with eight spatial basis functions.

\begin{figure}[!ht]
\center{\includegraphics[width=1.0\textwidth]{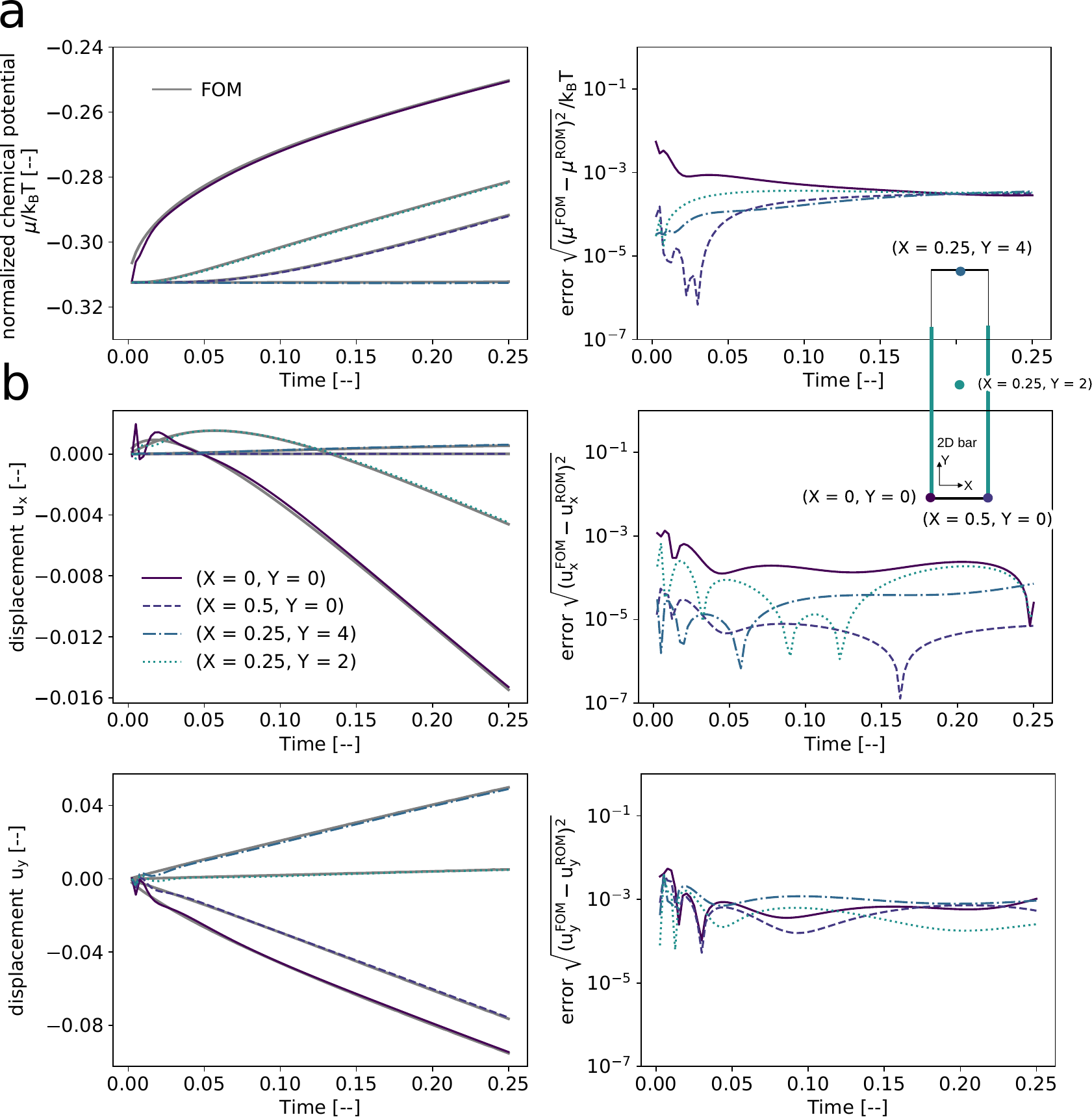}}
\caption{\textbf{Case study: hydrogel bar.} Nested-POD-based ROM simulation results with eight basis functions. \textbf{a.} normalized chemical potential and square root error with respect to the FOM; \textbf{b.} displacement magnitude and square root error with respect to the FOM.}
\label{fig:ROM_mu_disp_errors_gel_bar}
\end{figure}

The nested-POD-based ROM is assembled with six and eight basis functions for the temporal and spatial domains.
Four locations at the hydrogel bar domain are selected to verify the ROM’s accuracy as marked in Figure~\ref{fig:ROM_mu_disp_errors_gel_bar}. The root mean square error (RMSE) is measured over time at each location.

\begin{figure}[!ht]
\center{\includegraphics[width=1.0\textwidth]{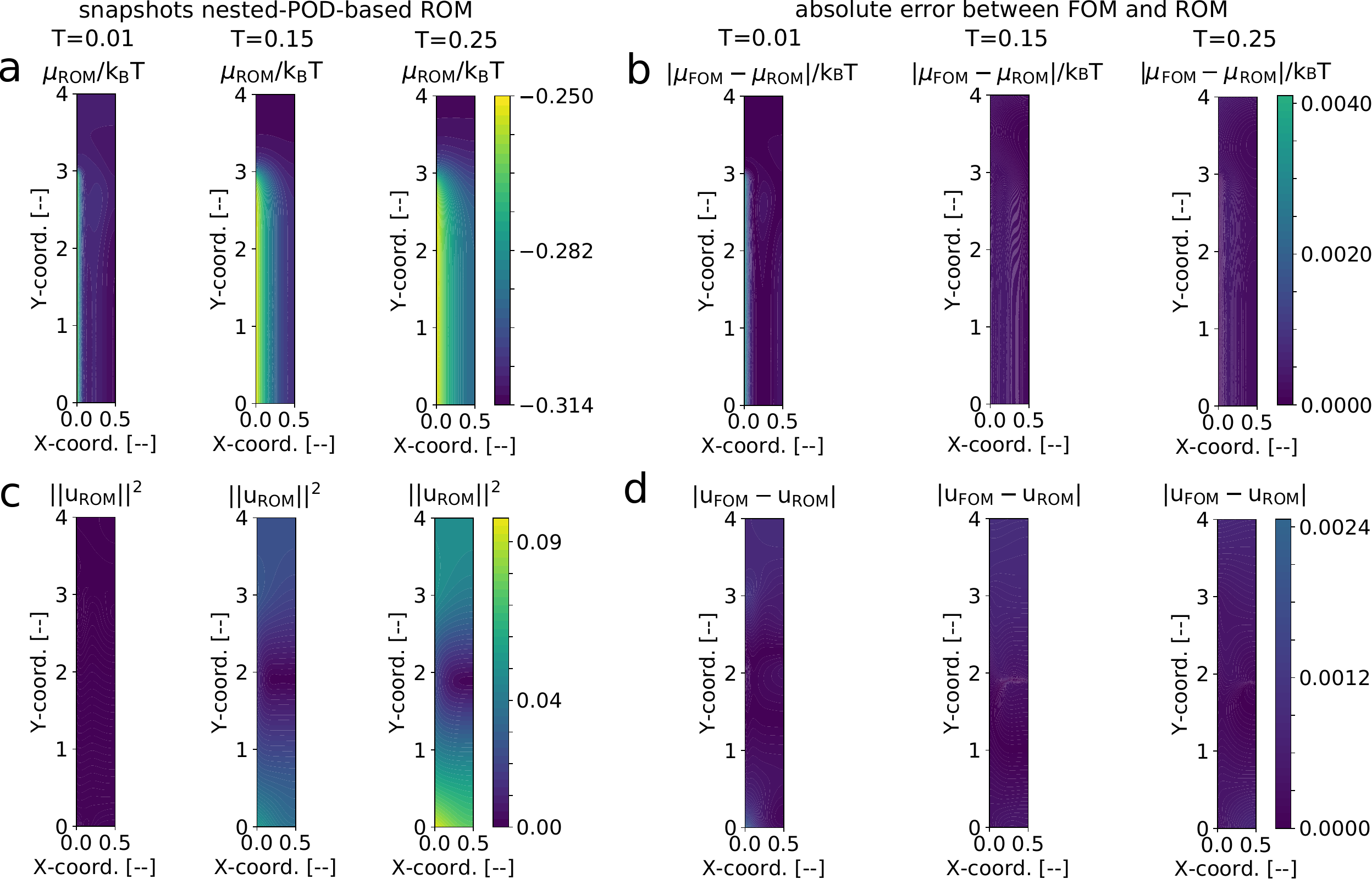}}
\caption{\textbf{Case study: hydrogel bar.} Nested-POD-based ROM simulation results. \textbf{a.} normalized chemical potential and \textbf{b.} absolute error of the ROM with respect to the FOM; \textbf{c.} displacement magnitude and \textbf{d.} absolute error of the ROM with respect to the FOM at time steps $T = \{0.01, 0.15, 0.25\}$}
\label{fig:ROM_mu_disp_gel_bar}
\end{figure}

The ROM for the hydrogel bar developed using the nested-POD method is compared with the nominal FOM model ($A=4000, \lambda=1558$ and $\alpha=0.66$). 
Figure~\ref{fig:ROM_mu_disp_errors_gel_bar} depicts a comparison of the temporal evolution between ROM and FOM in terms of chemical potential (Figure~\ref{fig:ROM_mu_disp_errors_gel_bar}\textbf{a}), displacement in $x$ direction and displacement in $y$ direction (Figure~\ref{fig:ROM_mu_disp_errors_gel_bar}\textbf{b}) at different points of the hydrogel bar. 
The figures show that the maximum RMSE between both models remains lower than $10^{-2}$ for the chemical potential and displacement. 

Figure~\ref{fig:ROM_mu_disp_gel_bar} displays the chemical potential (Figure~\ref{fig:ROM_mu_disp_gel_bar}\textbf{a}) and displacement (Figure~\ref{fig:ROM_mu_disp_gel_bar}\textbf{c}) obtained with eight basis functions for the temporal and spatial domains at three different time steps, namely, $T= \{0.01, 0.12, 0.25\}$. The corresponding errors with respect to the FOM are illustrated for the chemical potential in Figure~\ref{fig:ROM_mu_disp_gel_bar}\textbf{b} and for the displacement in Figure~\ref{fig:ROM_mu_disp_gel_bar}\textbf{d}, respectively. 

As seen from Figures~\ref{fig:ROM_mu_disp_errors_gel_bar} and \ref{fig:ROM_mu_disp_gel_bar}, the nested-POD-based ROM can predict the FOM to a satisfactory extent. From Figure~\ref{fig:ROM_mu_disp_gel_bar}, it can be observed that the maximum discrepancy between the FOM and ROM is less than 0.40\% for the chemical potential and 0.24\% for the displacement field, respectively. 

Notice that the displacement field is not symmetric. The asymmetry in the displacement field suggests uneven swelling of the hydrogel bar during the printing process. This is relevant because this uneven swelling could impact the mechanical integrity of the final printed structure, particularly in complex multi-layered scaffold designs, where stability is crucial. 

\subsubsection{Uncertainty propagation of the material parameter}
\label{subsec:unc_prop}

In this subsection, we exploit the ROM's efficiency to understand the effect of material parameters on two QoIs. 
We consider $\lambda$ and $A$ as random variables and propagate their uncertainty onto \textbf{i.} the chemical potential at the nozzle's tip and \textbf{ii.} the induced stresses within the domain due to swelling. 

\begin{figure}[!h]
\center{\includegraphics[width=0.95\textwidth]{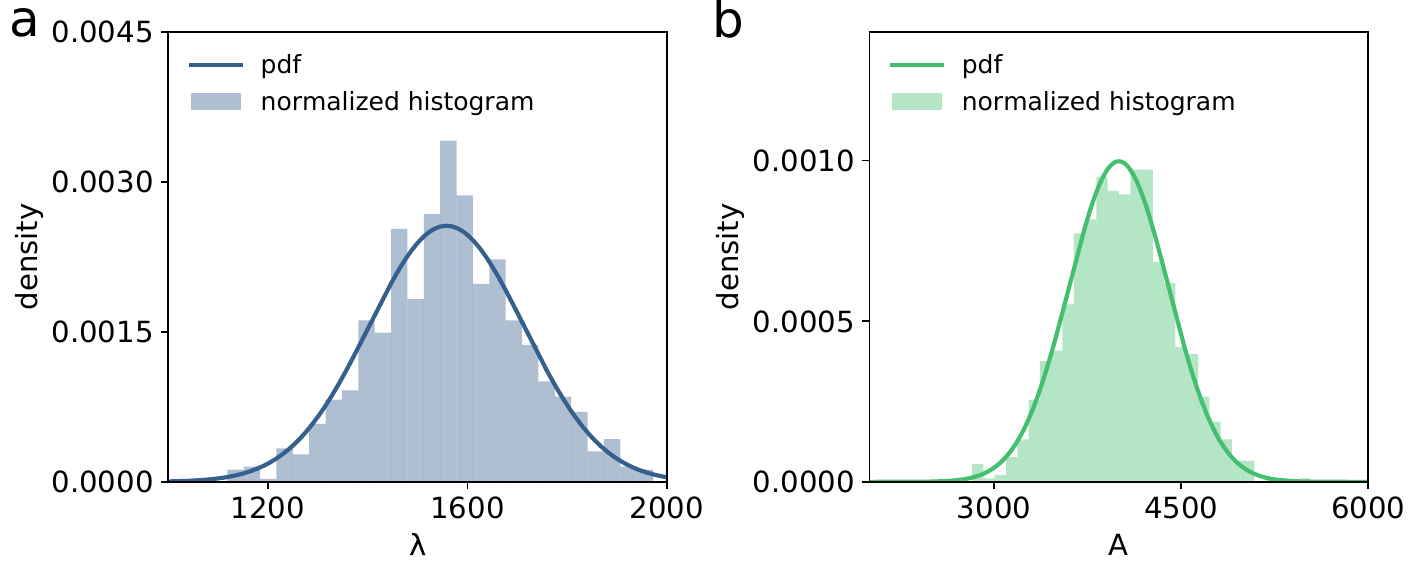}}
\caption{\textbf{Case study: hydrogel bar.} Uncertainty propagation analysis. Probability distribution function and normalized histogram of the generated samples \textbf{a.} of $\lambda$ and of $A$.}
\label{fig:gel_bar_UQ_pdfs_input}
\end{figure}

In principle, the probability density function (PDF) of the two material parameters can be determined as discussed below:
\begin{enumerate}
    \item \textbf{Bayesian updating} of the material parameters from measurement data; the reader can, for instance, refer to \cite{Noii2022BMU}, \cite{Grashorn2023TMaps}, and \cite{Anton2024}. However, acquiring species concentration-related and displacement measurement data during printing is a technological challenge. 
    \item \textbf{Bayesian inference from benchmark:}  Alternatively, one could quantify the uncertainty for the benchmark problem and then propagate this uncertainty in the case study. 
    However, although both problems involve the same hydrogel material, the distinct boundary conditions and domain dimensions result in different chemical potential profiles and displacement magnitudes, thereby influencing the material response differently, as evidenced when comparing Figures~\ref{fig:ROM_mu_disp_errors_gel_bar}, \ref{fig:ROM_mu_disp_gel_bar} (case study) with Figures~\ref{fig:FOM_mu_disp}, \ref{fig:FOM_time} (benchmark). Consequently, directly applying the identified PDFs from the benchmark problem may not adequately capture the full range of uncertainties present in the case study.
    \item \textbf{Designing synthetic PDFs based on prior knowledge or expert opinion} for $\lambda$ and $A$ around the nominal values identified for the benchmark problem is a more appropriate approach in the present context. Here we assume a normal distribution and use the Monte Carlo method to generate $1000$ samples for $\lambda$ and $A$. We set the mean values of $\lambda$ and $A$ equal to their nominal values $[\lambda, A] = [1558, 4000]$ and the standard deviation as $10\%$ around these nominal values. 
    The motivation for such a choice is based on the experimental work by \cite{Chen2020LinearHydrogelsExp} where it is evidenced that, under a similar hydrogel fabrication protocol, $\lambda$ and $A$ present a significant level of uncertainty (the reader is referred to Figures 3 and 5 in \cite{Chen2020LinearHydrogelsExp}).
    The reader should notice that assuming a Gaussian distribution implies that $\lambda$ and $A$ could take unphysical negative values, but the probability of getting them at the given mean and standard deviation is approximately $7 \times 10^{-24}$, i.e., essentially zero.
    Figure~\ref{fig:gel_bar_UQ_pdfs_input} depicts the probability density function (PDF) of along with the normalized histogram of $1000$ samples generated from the PDF for $\lambda$ (Figure~\ref{fig:gel_bar_UQ_pdfs_input}\textbf{a}) and $A$ (Figure~\ref{fig:gel_bar_UQ_pdfs_input}\textbf{b}). It is observed from the figure that, indeed, the two random variables follow a normal distribution, and the generated samples remain within the range the ROM was trained ($\lambda = [1000, 2000]$ and $A = [2000, 6000]$).
\end{enumerate}   

\begin{figure}[H]
\center{\includegraphics[width=0.95\textwidth]{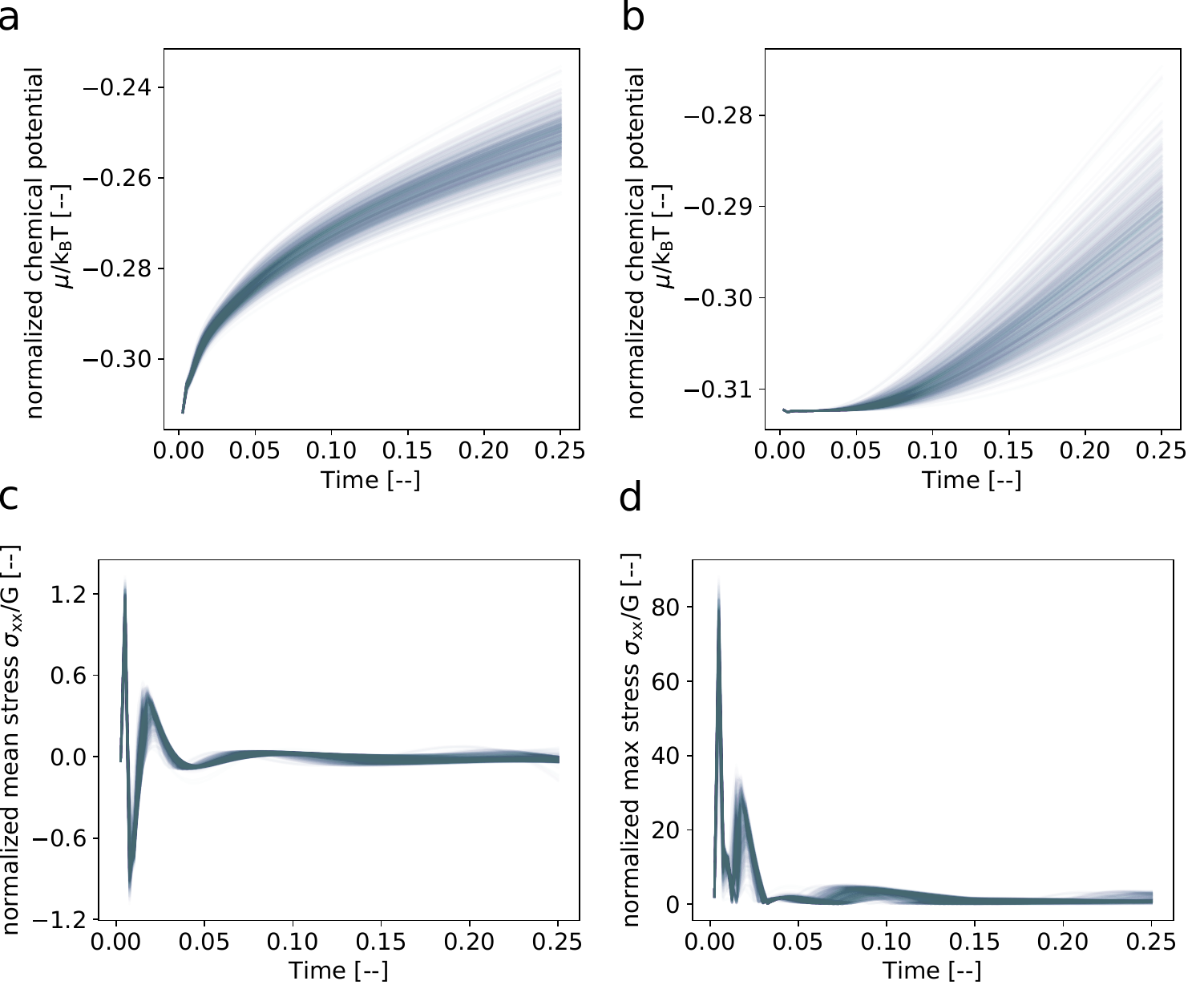}}
\caption{\textbf{Case study: hydrogel bar.} Uncertainty propagation analysis. Normalized chemical potential at the nozzle's tip at points \textbf{a.} $(X = 0, Y = 0)$ and \textbf{b.} $(X = 0.5, Y = 0)$. Normalized induced \textbf{c.} mean and \textbf{d.} maximum stress.}
\label{fig:gel_bar_UQ}
\end{figure}

We solve the nested-POD-based ROM iteratively and observe how the uncertainty in $\lambda$ and $A$ propagates to the QoIs. Figure~\ref{fig:gel_bar_UQ} shows the effect of the uncertain material parameters on the temporal evolution of the selected QoIs for $1000$ ROM realizations. It is worth remarking that one iteration of the ROM took an average of 15 seconds, which led to a total computation time of around 4 hours on a standard laptop. The same $1000$ iterations adopting the FOM instead will take up to 24 hours on the same laptop (an average of 90 seconds for each FOM iteration), which shows the advantage of adopting the nested-POD-based ROM in this many-query task. 

Figures~\ref{fig:gel_bar_UQ}\textbf{a} and \textbf{b} report the evolution of the normalized chemical potential at the nozzle's tip. Figure~\ref{fig:gel_bar_UQ}\textbf{a} refers to the corner at ($X=0,~Y=0$) and Figure~\ref{fig:gel_bar_UQ}\textbf{b} to the nozzle's tip center at the point ($X=0.5,~Y=0$).
From Figure~\ref{fig:gel_bar_UQ} can be observed that variations up to $10\%$ in the material parameters can lead to considerable differences in the diffusion-deformation process evolution. The effect of varying $\lambda$ and $A$ becomes more pronounced at the center of the nozzle's tip. 

The normalized mean induced stress in the \(XX\)-direction at each time instance is computed as the average of the stress values over the domain as
\begin{equation}
\sigma^{\text{mean}}_{XX}(t) = \frac{1}{A_{2D-block}} \int_{A_{2D-block}} \sigma_{xx}((x,y),t) dA,
\end{equation}
with $A_{2D-block}$ the simulated hydrogel bar area and $dA = dxdy$ representing the differential area element.

The maximum normalized induced stress in the \(XX\)-direction at each time instance is computed as the maximum value of the stress over the domain as
\begin{equation}
\sigma^{\text{max}}_{XX}(t) = \max_{(x, y) \in A_{\text{2D-block}}} \sigma_{XX}((x, y), t).
\end{equation}

Similarly, we can compute the normalized mean $\sigma^{\text{mean}}_{YY}$ and max $\sigma^{\text{max}}_{YY}$ induced stresses in the \(YY\)-direction at each time instance.

Figures~\ref{fig:gel_bar_UQ}\textbf{c} and \textbf{d} show the normalized mean $\sigma^{\text{mean}}_{XX}$ and maximum $\sigma^{\text{max}}_{XX}$ induced stresses in the \(XX\)-direction due to the diffusion of the solvent into the hydrogel, but identical results are obtained for $\sigma^{\text{mean}}_{YY}$ and $\sigma^{\text{max}}_{YY}$. It is observed from the figures that the level of induced stress has pronounced peaks at the beginning of the simulation, which can lead to some damage to the cells contained in bio-inks.

These induced stresses are normalized by the shear modulus $G$, which \cite{Chen2020LinearHydrogelsExp} found to be equal to $33$ kPa from the experiments. Hence, the maximum induced stress can be amplified $80$ times and reaches a maximum of $2600$ kPa based on the results reported in Figure~\ref{fig:gel_bar_UQ}\textbf{d}. Just as a reference example, \cite{Chirianni2024BioprintingStress} studied the level of damage caused by the shear stresses during hydrogels extrusion in bioprinting applications and reported that cells could suffer damage up to $50\%$ for induced stresses between $100$ and $500$ kPa. This threshold is clearly exceeded at the beginning of the performed simulation and can have significant consequences on the cell viability at the end of the process.
One alternative to preventing such high-stress fluctuations could be implementing a feedback control mechanism that can monitor the mean and maximum induced stresses informed by the ROM and dynamically adjust the solvent concentration at the mixing zone (see Figure~\ref{fig:simulation_setup}\textbf{b}) to keep the induced stresses under a desired limit. 

\section{Conclusions}
\label{sec:conclusions}

In this paper, we demonstrated that POD-based algorithms are feasible for developing a ROM of a coupled diffusion-deformation model. The accuracy of the ROM was tested in two different scenarios: \textbf{i.} free-swelling of a 2D gel block, a well-established benchmark in the literature, and \textbf{ii.} a case study resembling a bioprinting process using a co-axial bioprinter. 
In both cases, we highlighted the computational benefits of adopting the ROM, where the computational cost of the ROM online evaluation was only 0.01\% of running the full-order model. Furthermore, when using a finer mesh, we found that the ROM is even $5172$ times faster than the FOM in simulating the time-dependent diffusion-deformation evolution in the co-axial printing application.

We demonstrated the benefits of adopting a ROM in many-query tasks, that is, parameter identification from full-field data and uncertainty propagation.
In parameter identification, we showed that the computational resources spent in the ofﬂine phase for training data generation and ROM construction are repaid by a relatively large number of evaluations (90 on average) of the trained model in an online phase.
Additionally, we exploited the simulation speed-up provided by the ROM and propagated the uncertainty in the material parameters to two quantities of interest in co-axial bioprinting: the chemical potential at the nozzle's tip and the stresses due to the fluid's absorption that lead to the hydrogel's swelling.
The significant simulation speed-up provided by the ROM paves the way for more advanced applications, such as online feedback error compensation during the bioprinting process. This would ensure that the printed products meet design and functional requirements, ultimately aiming to produce consistent, high-quality hydrogel products to advance the reliability of bioprinting technologies. 

\backmatter

\section*{Declarations}

\bmhead{Supplementary information}
No supplementary information was produced from this research.

\bmhead{Acknowledgments}
GA gratefully acknowledges Dr. Francesco Ballarin, the RBniCS developer, for his support in the implementation of the reduced-order model.

\textbf{Funding:}
HW acknowledges the DFG for funding within the individual research grant with project ID 512730472.
TW gratefully acknowledges the Deutsche Forschungsgemeinschaft (DFG) under Germany’s Excellence Strategy within the Cluster of Excellence PhoenixD (EXC 2122, Project ID 390833453).

\noindent\textbf{Conflict of interest/Competing interests:}
On behalf of all authors, the corresponding author states that there is no conflict of interest.

\noindent\textbf{Code availability:}
The code to reproduce the results presented in this manuscript can be accessed in the following Zenodo repository: \url{https://doi.org/10.5281/zenodo.11356177}.


\setlength{\bibsep}{5pt} 
\bibliography{sn-bibliography}

\begin{thebibliography}{64}
\providecommand{\natexlab}[1]{#1}
\providecommand{\url}[1]{{#1}}
\providecommand{\urlprefix}{URL }
\providecommand{\doi}[1]{\url{https://doi.org/#1}}
\providecommand{\eprint}[2][]{\url{#2}}
 \bibcommenthead

\bibitem[{Agarwal et~al(2024)Agarwal, Urrea-Quintero, Wessels, and Wick}]{Urrea2024MOR_hydrogels_code}
Agarwal G, Urrea-Quintero JH, Wessels H, et~al (2024) Code: {P}arameter identification and uncertainty propagation of hydrogel coupled diffusion-deformation using {POD-based} reduced-order modeling. \urlprefix\url{https://doi.org/10.5281/zenodo.11356177}

\bibitem[{Aln{\ae}s et~al(2015)Aln{\ae}s, Blechta, Hake, Johansson, Kehlet, Logg, Richardson, Ring, Rognes, and Wells}]{Alnaes2015Fenics}
Aln{\ae}s M, Blechta J, Hake J, et~al (2015) {The {FEniCS} project version 1.5}. Archive of Numerical Software 3(100)

\bibitem[{Anand(2015)}]{Anand2015LinearPoroelasticity}
Anand L (2015) {2014 Drucker medal paper: A derivation of the theory of linear poroelasticity from chemoelasticity}. Journal of Applied Mechanics 82(11):111005

\bibitem[{Anand and Govindjee(2020)}]{Anand2020CMM_solids_Book}
Anand L, Govindjee S (2020) Continuum mechanics of solids. Oxford University Press

\bibitem[{Anton et~al(2024)Anton, Tröger, Wessels, Römer, Henkes, and Hartmann}]{Anton2024}
Anton D, Tröger JA, Wessels H, et~al (2024) Deterministic and statistical calibration of constitutive models from full-field data with parametric physics-informed neural networks. \eprint{arXiv:2405.18311 [cs.LG]}

\bibitem[{Ballarin et~al(2015)Ballarin, Manzoni, Quarteroni, and Rozza}]{ballarin_supremizer_2015}
Ballarin F, Manzoni A, Quarteroni A, et~al (2015) Supremizer stabilization of {POD}–{Galerkin} approximation of parametrized steady incompressible {Navier}–{Stokes} equations. Int J Numer Methods Eng 102(5):1136--1161

\bibitem[{Ballarin et~al(2022)Ballarin, Rozza, and Strazzullo}]{Ballarin2022spaceTimePODControl}
Ballarin F, Rozza G, Strazzullo M (2022) Space-time {POD-Galerkin} approach for parametric flow control. In: Handbook of Numerical Analysis, vol~23. Elsevier, p 307--338

\bibitem[{Benner et~al(2015{\natexlab{a}})Benner, Cohen, Ohlberger, and Willcox}]{BeCoOhlWill15}
Benner P, Cohen A, Ohlberger M, et~al (2015{\natexlab{a}}) Model Reduction and Approximation: Theory and Algorithms. SIAM Philadelphia

\bibitem[{Benner et~al(2015{\natexlab{b}})Benner, Gugercin, and Willcox}]{Benner2015surveyPROM}
Benner P, Gugercin S, Willcox K (2015{\natexlab{b}}) A survey of projection-based model reduction methods for parametric dynamical systems. SIAM review 57(4):483--531

\bibitem[{Benner et~al(2020)Benner, Schilders, Grivet-Talocia, Quarteroni, Rozza, and Miguel~Silveira}]{benner2020model}
Benner P, Schilders W, Grivet-Talocia S, et~al (2020) {M}odel {O}rder {R}eduction: Volume 2: Snapshot-Based Methods and Algorithms. De Gruyter

\bibitem[{Biot(1941)}]{Biot1941general}
Biot MA (1941) General theory of three-dimensional consolidation. Journal of applied physics 12(2):155--164

\bibitem[{Bonatti et~al(2021)Bonatti, Chiesa, Vozzi, and De~Maria}]{Bonatti2021SimBioprinting}
Bonatti AF, Chiesa I, Vozzi G, et~al (2021) Open-source {CAD-CAM} simulator of the extrusion-based bioprinting process. Bioprinting 24:e00172

\bibitem[{Bouklas et~al(2015)Bouklas, Landis, and Huang}]{Bouklas2015Nonlinear}
Bouklas N, Landis CM, Huang R (2015) A nonlinear, transient finite element method for coupled solvent diffusion and large deformation of hydrogels. Journal of the Mechanics and Physics of Solids 79:21--43

\bibitem[{Brand(2006)}]{Brand2006IncrementalSVD}
Brand M (2006) Fast low-rank modifications of the thin singular value decomposition. Linear algebra and its applications 415(1):20--30

\bibitem[{Brunton and Noack(2015)}]{Brunton2015CLturbolence}
Brunton SL, Noack BR (2015) Closed-loop turbulence control: Progress and challenges. Applied Mechanics Reviews 67(5):050801

\bibitem[{Caccavo et~al(2018)Caccavo, Cascone, Lamberti, and Barba}]{Caccavo2018hydrogels}
Caccavo D, Cascone S, Lamberti G, et~al (2018) Hydrogels: experimental characterization and mathematical modelling of their mechanical and diffusive behaviour. Chemical Society Reviews 47(7):2357--2373

\bibitem[{Chen et~al(2015)Chen, Quarteroni, and Rozza}]{Chen2015ROMUQ}
Chen P, Quarteroni A, Rozza G (2015) Reduced order methods for uncertainty quantification problems. ETH Zurich, SAM Report 3

\bibitem[{Chen et~al(2020)Chen, Huang, and Ravi-Chandar}]{Chen2020LinearHydrogelsExp}
Chen S, Huang R, Ravi-Chandar K (2020) Linear and nonlinear poroelastic analysis of swelling and drying behavior of gelatin-based hydrogels. International Journal of Solids and Structures 195:43--56

\bibitem[{Chester and Anand(2010)}]{Chester2010DiffDeform}
Chester SA, Anand L (2010) A coupled theory of fluid permeation and large deformations for elastomeric materials. Journal of the Mechanics and Physics of Solids 58(11):1879--1906

\bibitem[{Chester and Anand(2011)}]{Chester2011thermo}
Chester SA, Anand L (2011) A thermo-mechanically coupled theory for fluid permeation in elastomeric materials: application to thermally responsive gels. Journal of the Mechanics and Physics of Solids 59(10):1978--2006

\bibitem[{Chester et~al(2015)Chester, Di~Leo, and Anand}]{Chester2015Abaqus}
Chester SA, Di~Leo CV, Anand L (2015) A finite element implementation of a coupled diffusion-deformation theory for elastomeric gels. International Journal of Solids and Structures 52:1--18

\bibitem[{Chirianni et~al(2024)Chirianni, Vairo, and Marino}]{Chirianni2024BioprintingStress}
Chirianni F, Vairo G, Marino M (2024) Development of process design tools for extrusion-based bioprinting: From numerical simulations to nomograms through reduced-order modeling. Computer Methods in Applied Mechanics and Engineering 419:116685

\bibitem[{Eftekhar~Azam et~al(2017)Eftekhar~Azam, Mariani, and Attari}]{Eftekhar2017DamageDetection}
Eftekhar~Azam S, Mariani S, Attari N (2017) Online damage detection via a synergy of proper orthogonal decomposition and recursive bayesian filters. Nonlinear Dynamics 89:1489--1511

\bibitem[{Fareed et~al(2018)Fareed, Singler, Zhang, and Shen}]{Fareed2018iPOD}
Fareed H, Singler JR, Zhang Y, et~al (2018) Incremental proper orthogonal decomposition for {PDE} simulation data. Computers \& Mathematics with Applications 75(6):1942--1960

\bibitem[{Fischer et~al(2024)Fischer, Roth, Wick, Chamoin, and Fau}]{Fischer2024iPOD}
Fischer H, Roth J, Wick T, et~al (2024) {MORe DWR:} space-time goal-oriented error control for incremental {POD-based ROM} for time-averaged goal functionals. Journal of Computational Physics p 112863

\bibitem[{Girfoglio et~al(2022)Girfoglio, Quaini, and Rozza}]{GIRFOGLIO2022105536}
Girfoglio M, Quaini A, Rozza G (2022) {A {POD-Galerkin} reduced order model for the Navier–Stokes equations in stream function-vorticity formulation}. Comp Fluids 244:105536

\bibitem[{Grashorn et~al(2023)Grashorn, Urrea-Quintero, Broggi, Chamoin, and Beer}]{Grashorn2023TMaps}
Grashorn J, Urrea-Quintero JH, Broggi M, et~al (2023) Transport map bayesian parameter estimation for dynamical systems. PAMM 23(1):e202200136

\bibitem[{Gr{\"a}{\ss}le and Hinze(2018)}]{grassle2018pod}
Gr{\"a}{\ss}le C, Hinze M (2018) {POD} reduced-order modeling for evolution equations utilizing arbitrary finite element discretizations. Adv Comput Math 44(6):1941--1978

\bibitem[{Haasdonk et~al(2023)Haasdonk, Kleikamp, Ohlberger, Schindler, and Wenzel}]{Bernard2022}
Haasdonk B, Kleikamp H, Ohlberger M, et~al (2023) A new certified hierarchical and adaptive rb-ml-rom surrogate model for parametrized pdes. SIAM Journal on Scientific Computing 45(3):A1039--A1065

\bibitem[{Hajikhani et~al(2021)Hajikhani, Wriggers, and Marino}]{Hajikhani2021chemomechanics}
Hajikhani A, Wriggers P, Marino M (2021) Chemo-mechanical modelling of swelling and crosslinking reaction kinetics in alginate hydrogels: A novel theory and its numerical implementation. Journal of the Mechanics and Physics of Solids 153:104476

\bibitem[{Hemati et~al(2014)Hemati, Williams, and Rowley}]{Hemati2014StreamingSVD}
Hemati MS, Williams MO, Rowley CW (2014) Dynamic mode decomposition for large and streaming datasets. Physics of Fluids 26(11)

\bibitem[{Irick and Brown(2019)}]{Kevin2019}
Irick K, Brown E (2019) In-situ thermal {ROM}-based optimization using {Borg MOEA}: A preliminary study. In: Heat Transfer Summer Conference, American Society of Mechanical Engineers, p V001T02A005

\bibitem[{Kadeethum et~al(2021)Kadeethum, Ballarin, and Bouklas}]{Kadeethum2021DDMOR}
Kadeethum T, Ballarin F, Bouklas N (2021) Data-driven reduced order modeling of poroelasticity of heterogeneous media based on a discontinuous {Galerkin} approximation. GEM-International Journal on Geomathematics 12(1):12

\bibitem[{Kadeethum et~al(2022)Kadeethum, Ballarin, Choi, O’Malley, Yoon, and Bouklas}]{Kadeethum2022CAEN}
Kadeethum T, Ballarin F, Choi Y, et~al (2022) Non-intrusive reduced order modeling of natural convection in porous media using convolutional autoencoders: Comparison with linear subspace techniques. Advances in Water Resources 160:104098

\bibitem[{Kerschen et~al(2005)Kerschen, Golinval, Vakakis, and Bergman}]{Kerschen2005}
Kerschen G, Golinval JC, Vakakis AF, et~al (2005) The method of {P}roper {O}rthogonal {D}ecomposition for dynamical characterization and order reduction of mechanical systems: An overview. Nonlinear Dyn 41(1):147--169

\bibitem[{Kjar et~al(2021)Kjar, McFarland, Mecham, Harward, and Huang}]{Kjar2021coaxial_print}
Kjar A, McFarland B, Mecham K, et~al (2021) Engineering of tissue constructs using coaxial bioprinting. Bioactive materials 6(2):460--471

\bibitem[{K{\"u}hl et~al(2024)K{\"u}hl, Fischer, Hinze, and Rung}]{Kuhl2024IncrementalSVD}
K{\"u}hl N, Fischer H, Hinze M, et~al (2024) An incremental singular value decomposition approach for large-scale spatially parallel \& distributed but temporally serial data--applied to technical flows. Computer Physics Communications 296:109022

\bibitem[{Lange et~al(2024)Lange, H{\"u}tter, and Kiefer}]{Nils2024}
Lange N, H{\"u}tter G, Kiefer B (2024) A monolithic hyper rom fe2 method with clustered training at finite deformations. Computer Methods in Applied Mechanics and Engineering 418:116522

\bibitem[{Lass and Volkwein(2014)}]{Lass2014adaptivePOD}
Lass O, Volkwein S (2014) Adaptive {POD} basis computation for parametrized nonlinear systems using optimal snapshot location. Computational Optimization and Applications 58:645--677

\bibitem[{Lassila et~al(2014{\natexlab{a}})Lassila, Manzoni, Quarteroni, and Rozza}]{Lassila2014MOR}
Lassila T, Manzoni A, Quarteroni A, et~al (2014{\natexlab{a}}) Model order reduction in fluid dynamics: challenges and perspectives. Reduced Order Methods for modeling and computational reduction pp 235--273

\bibitem[{Lassila et~al(2014{\natexlab{b}})Lassila, Manzoni, Quarteroni, and Rozza}]{Lassila2014ROM}
Lassila T, Manzoni A, Quarteroni A, et~al (2014{\natexlab{b}}) Model order reduction in fluid dynamics: challenges and perspectives. Reduced Order Methods for modeling and computational reduction pp 235--273

\bibitem[{Liu et~al(2016)Liu, Zhang, Zhang, and Zheng}]{Liu2016TransGels}
Liu Y, Zhang H, Zhang J, et~al (2016) Transient swelling of polymeric hydrogels: A new finite element solution framework. International Journal of Solids and Structures 80:246--260

\bibitem[{Lu et~al(2015{\natexlab{a}})Lu, Yu, Chen, Cao, and Hou}]{Lu2015RotorSys}
Lu K, Yu H, Chen Y, et~al (2015{\natexlab{a}}) A modified nonlinear {POD} method for order reduction based on transient time series. Nonlinear Dynamics 79:1195--1206

\bibitem[{Lu et~al(2015{\natexlab{b}})Lu, Yu, Chen, Cao, and Hou}]{Lu2015TransientPOD}
Lu K, Yu H, Chen Y, et~al (2015{\natexlab{b}}) A modified nonlinear {POD} method for order reduction based on transient time series. Nonlinear Dynamics 79:1195--1206

\bibitem[{Lu et~al(2016)Lu, Chen, Jin, and Hou}]{Lu2016RotorSys}
Lu K, Chen Y, Jin Y, et~al (2016) Application of the transient proper orthogonal decomposition method for order reduction of rotor systems with faults. Nonlinear Dynamics 86:1913--1926

\bibitem[{Lu et~al(2019)Lu, Jin, Chen, Yang, Hou, Zhang, Li, and Fu}]{Lu2019PODreview}
Lu K, Jin Y, Chen Y, et~al (2019) Review for order reduction based on proper orthogonal decomposition and outlooks of applications in mechanical systems. Mechanical Systems and Signal Processing 123:264--297

\bibitem[{Martinsson et~al(2011)Martinsson, Rokhlin, and Tygert}]{Martinsson2011RandomizedSVD}
Martinsson PG, Rokhlin V, Tygert M (2011) A randomized algorithm for the decomposition of matrices. Applied and Computational Harmonic Analysis 30(1):47--68

\bibitem[{Negri et~al(2013)Negri, Rozza, Manzoni, and Quarteroni}]{Negri2013PODControl}
Negri F, Rozza G, Manzoni A, et~al (2013) Reduced basis method for parametrized elliptic optimal control problems. SIAM Journal on Scientific Computing 35(5):A2316--A2340

\bibitem[{Noii et~al(2022)Noii, Khodadadian, Ulloa, Aldakheel, Wick, Francois, and Wriggers}]{Noii2022BMU}
Noii N, Khodadadian A, Ulloa J, et~al (2022) Bayesian inversion with open-source codes for various one-dimensional model problems in computational mechanics. Archives of Computational Methods in Engineering 29(6):4285--4318

\bibitem[{Nonino et~al(2021)Nonino, Ballarin, and Rozza}]{nonino_monolithic_2021}
Nonino M, Ballarin F, Rozza G (2021) A monolithic and a partitioned, reduced basis method for fluid-structure interaction problems. Fluids 6(6):229

\bibitem[{Pagani and Manzoni(2021)}]{Pagani2021EnablingUQ_POD}
Pagani S, Manzoni A (2021) Enabling forward uncertainty quantification and sensitivity analysis in cardiac electrophysiology by reduced order modeling and machine learning. International Journal for Numerical Methods in Biomedical Engineering 37(6):e3450

\bibitem[{R{\"o}mer et~al(2024)R{\"o}mer, Hartmann, Tr{\"o}ger, Anton, Wessels, Flaschel, and De~Lorenzis}]{romer2024reduced}
R{\"o}mer U, Hartmann S, Tr{\"o}ger JA, et~al (2024) Reduced and all-at-once approaches for model calibration and discovery in computational solid mechanics. arXiv preprint arXiv:240416980

\bibitem[{Rozza et~al(2024)Rozza, Ballarin, Scandurra, and Pichi}]{Rozza2024RBniCsBook}
Rozza G, Ballarin F, Scandurra L, et~al (2024) Real Time Reduced Order Computational Mechanics. SISSA Springer Series, Springer Cham

\bibitem[{Sahyoun and Djouadi(2013)}]{Sahyoun2013localPOD}
Sahyoun S, Djouadi S (2013) Local proper orthogonal decomposition based on space vectors clustering. In: 3rd International Conference on Systems and Control, IEEE, pp 665--670

\bibitem[{Sirovich(1987)}]{Sirovich1987SnapshotsSVD}
Sirovich L (1987) Turbulence and the dynamics of coherent structures. i - iii. Quarterly of applied mathematics 45(3):561--571

\bibitem[{Strazzullo et~al(2020)Strazzullo, Zainib, Ballarin, and Rozza}]{Strazzullo2020ROMControl}
Strazzullo M, Zainib Z, Ballarin F, et~al (2020) Reduced order methods for parametrized non-linear and time-dependent optimal flow control problems, towards applications in biomedical and environmental sciences. In: Numerical Mathematics and Advanced Applications ENUMATH 2019: European Conference, Egmond aan Zee, The Netherlands, September 30-October 4, Springer, pp 841--850

\bibitem[{Urrea-Quintero et~al(2024)Urrea-Quintero, Marino, Wick, and Nackenhorst}]{Urrea2023DiffDef_hydrogels}
Urrea-Quintero JH, Marino M, Wick T, et~al (2024) A comparative analysis of transient finite-strain coupled diffusion-deformation theories for hydrogels. Archives of Computational Methods in Engineering --:50

\bibitem[{Vaccaro(1991)}]{Vaccaro1991SVD_signal}
Vaccaro RJ (1991) {SVD} and signal processing, ii. algorithms, analysis and applications. Amsterdam: Elsevier

\bibitem[{Virtanen et~al(2020)Virtanen, Gommers, Oliphant, Haberland, Reddy, Cournapeau, Burovski, Peterson, Weckesser, Bright et~al}]{Virtanen2020scipy}
Virtanen P, Gommers R, Oliphant TE, et~al (2020) {SciPy 1.0}: fundamental algorithms for scientific computing in {Python}. Nature methods 17(3):261--272

\bibitem[{Volkwein(2001)}]{Volkwein2001}
Volkwein S (2001) Optimal control of a phase-field model using {P}roper {O}rthogonal {D}ecomposition. ZAMM - Journal of Applied Mathematics and Mechanics / Zeitschrift für Angewandte Mathematik und Mechanik 81(2):83--97

\bibitem[{Wloka(1987)}]{Wlo87}
Wloka J (1987) Partial differential equations. Cambridge University Press

\bibitem[{Zhang and Khademhosseini(2017)}]{Zhang2017HydrogelsScience}
Zhang YS, Khademhosseini A (2017) Advances in engineering hydrogels. Science 356(6337):eaaf3627

\bibitem[{Zhu et~al(1997)Zhu, Byrd, Lu, and Nocedal}]{Zhu1997LBFGSB}
Zhu C, Byrd RH, Lu P, et~al (1997) Algorithm 778: {L-BFGS-B}: Fortran subroutines for large-scale bound-constrained optimization. ACM Transactions on mathematical software (TOMS) 23(4):550--560

\bibitem[{Zou et~al(2018)Zou, Conti, D{\'\i}ez, and Auricchio}]{Zou2018PGD_Ident_Biomechs}
Zou X, Conti M, D{\'\i}ez P, et~al (2018) A nonintrusive proper generalized decomposition scheme with application in biomechanics. International Journal for Numerical Methods in Engineering 113(2):230--251

\end{thebibliography}

\newpage
\begin{appendices}

\section{Thermodynamically consistent constitutive theory}
\label{app:const_theory}

In this Appendix, following standard thermodynamic arguments, we derive a thermodynamically consistent constitutive theory describing the coupled diffusion-deformation mechanisms for linear isotropic gels. First, the constitutive model is introduced in a general form. Then, the specific form of the constitutive equations for a previously swollen hydrogel is presented.

\subsection{Constitutive model and thermodynamic restrictions}

The local form of the second law of thermodynamics that links fluid's concentration changes due to diffusion inside a previously swollen hydrogel with the hydrogel's deformation due to swelling reads
\begin{equation}\label{eq:secondlaw-0}
    \bm{\sigma} : \dot{\bm{\varepsilon}} + \mu \dot{c} - \mathbf{j} \cdot \text{grad}(\mu) - \dot{\psi} \geq 0,
\end{equation}
where $\psi:\mathcal{B}\times I\to\mathbb{R}$ is the free energy density function. Guided by equation \eqref{eq:secondlaw-0}, the following constitutive response functions for the free energy $\psi$, the Cauchy stress $\bm{\sigma}$, and the chemical potential $\mu$ are expressed in terms of fluid concentration $c$ and  strain  $\bm{\varepsilon}$ as 
\begin{equation} \label{eq:PsiR}
\psi = \hat{\psi}(c, \bm{\varepsilon}), ~~ \bm{\sigma} = \hat{\bm{\sigma}}(c, \bm{\varepsilon}), ~~ \text{and} ~~ \mu = \hat{\mu}(c, \bm{\varepsilon}).
\end{equation}

The reader should recall that, based on Fick's law for the species diffusion, the fluid flux $\mathbf{j}$ is given by equation~\eqref{eq:flux_j}.

Therefore, equation \eqref{eq:secondlaw-0} can be reformulated as
\begin{equation}\label{eq:secondlaw}
    \left( \dfrac{\partial \hat{\psi} (c, \bm{\varepsilon})}{\partial \bm{\varepsilon}} -  \bm{\sigma} \right) : \dot{\bm{\varepsilon}} + \left( \dfrac{\partial \hat{\psi} (c, \bm{\varepsilon})}{\partial c} - \mu \right) \dot{c} -  \text{grad}(\mu) \cdot \mathbf{M}~ \text{grad}(\mu) \leq 0,
\end{equation}
for all $\bm{\sigma}$, $\bm{\varepsilon}$, $c$, and $\mu$ fields. 
Therefore, the following thermodynamically-consistent constitutive relations can be established for Cauchy's stress
\begin{equation}\label{eq:gen_sigma}
    \hat{\bm{\sigma}}(c, \bm{\varepsilon}) = \dfrac{\partial \hat{\psi} (c, \bm{\varepsilon})}{\partial \bm{\varepsilon}},
\end{equation}
and for the chemical potential
\begin{equation}\label{eq:gen_mu}
    \hat{\mu}(c, \bm{\varepsilon}) = \dfrac{\partial \hat{\psi} (c, \bm{\varepsilon})}{\partial c}.
\end{equation}
Additionally, since
\begin{equation}
    \text{grad}(\mu) \cdot \mathbf{M}~ \text{grad}(\mu) \geq 0,
\end{equation}
for all $\bm{\varepsilon}$, $c$, and $\mu$, thus, $\mathbf{M}$ has to be positive semi-definite. 

\subsubsection{Chemical potential as independent variable}

The strain $\bm{\varepsilon}$ and the concentration $c$ are the natural choice of independent variables for problems involving little or no species diffusion. However, for processes in which species diffusion is important, as in the hydrogels case, replacing constitutive dependence upon $c$ with constitutive dependence upon the chemical potential $\mu$ is preferable.

Assuming that $\partial \hat{\mu}(c, \bm{\varepsilon}) / \partial c > 0$ and continuously differentiable, it can be proved that for each fixed $\bm{\varepsilon}$ the relation $\mu = \hat{\mu}(c, \bm{\varepsilon})$ is invertible in $c$, so that $c = \breve{c}(\mu, \bm{\varepsilon})$.
Thus, bearing in mind that a  $\breve{\bullet}$ ``breve'' denotes a function of $(\mu,\bm{\varepsilon})$ while a  $\hat{\bullet}$ ``hat'' denotes a function of $(c,\bm{\varepsilon})$, the following grand-canonical energy function can be defined \citep{Anand2020CMM_solids_Book} [Chapter 15.5.3]
\begin{equation}\label{eq:gen_omega}
    \omega = \breve{\omega}(\mu, \bm{\varepsilon}) = \hat{\psi}(\breve{c}(\mu, \bm{\varepsilon}), \bm{\varepsilon}) - \breve{c}(\mu, \bm{\varepsilon}) \mu.
\end{equation}
Consequently, equivalent constitutive relations to those given by equations \eqref{eq:gen_sigma} and \eqref{eq:gen_mu} can be established for the stress 
\begin{equation}\label{eq:gen_sigma_breve}
    \breve{\bm{\sigma}}(\mu, \bm{\varepsilon}) = \dfrac{\partial \breve{\omega}(\mu, \bm{\varepsilon})}{\partial \bm{\varepsilon}}
\end{equation}
and for the fluid concentration
\begin{equation}\label{eq:gen_c_breve}
    \breve{c}(\mu, \bm{\varepsilon}) = - \dfrac{\partial \breve{\omega}(\mu, \bm{\varepsilon})}{\partial \mu}.
\end{equation}

\subsubsection{Specialization of the constitutive theory for linear isotropic gels}

If the material is isotropic, then $\breve{\omega}(\mu, \bm{\varepsilon})$ takes the following specific form \citep{Anand2020CMM_solids_Book}[Chapter 15.5.3]\footnote{The reader is referred to Chapter 15 by \cite{Anand2020CMM_solids_Book} for more details regarding the interpretation of the specific form of the grand-canonical free energy function and the material parameters.}
\begin{equation}\label{eq:specific_omega}
    \breve{\omega}(\mu, \bm{\varepsilon}) = - c_{0} (\mu - \mu_{0}) + G\vert \bm{\varepsilon} \vert^{2} + \frac{1}{2} \left( K^{d} - \frac{2}{3}G \right)(\text{tr} \bm{\varepsilon})^{2} + \frac{\chi}{\Lambda}(\mu - \mu_{0}) \text{tr} \bm{\varepsilon} - \frac{1}{2} \frac{1}{\Lambda}(\mu - \mu_{0})^2,   
\end{equation}
where ${\mu}_{0}$ is the initial chemical potential which defines the chemical potential at a relaxed state when the internal stresses are zero. $G$ and $K^{d}$ are the elastic shear and drained bulk modulus, respectively. 
$\chi$ is the stress-chemical modulus and $\Lambda$ the chemical modulus.

Therefore, $\breve{\bm{\sigma}}(\mu, \bm{\varepsilon})$ (equation~\eqref{eq:gen_sigma_breve}) is obtained as in equation~\eqref{eq:specific_sigma} and $\breve{c}(\mu, \bm{\varepsilon})$ (equation~\eqref{eq:gen_c_breve}) takes the form of equation~\eqref{eq:specific_mu}. The reader should notice that the  ``breve'' ($\breve{\bullet}$) is committed in equations~\eqref{eq:specific_sigma} and \eqref{eq:specific_mu} for simplicity in the notation.

This constitutive theory can be linked to Biot's classical theory of linear isotropic poroelasticity \citep{Biot1941general}. In that case, the stress chemical modulus $\chi$ and the chemical modulus $\Lambda$ are expressed as
\begin{equation}
    \chi = -\alpha M \Omega, ~~ \text{and} ~~ \Lambda = M \Omega^2,
\end{equation}
where $M$ is a Biot modulus, $\alpha$ is a Biot effective stress coefficient and $\Omega$ is the molar volume of the fluid with units of m$^3$/mol. 

In hydrogels, it is typically assumed that volume changes are solely due to variations in fluid concentration, considering the fluid molecules as incompressible with a constant molar volume $\Omega$. This assumption is formalized by setting $\alpha \to 1$ and $M \to \infty$, leading to $\Lambda \to \infty$ and $\chi/\Lambda \to -1/\Omega$. Hence, the last term of equation \eqref{eq:specific_mu} vanishes, resulting in the following kinematic constraint
\begin{equation}\label{eq:kinematic_cons}
    \Omega \left( \breve{c} - c_0 \right) = \text{tr} \bm{\varepsilon},
\end{equation}
indicating that the gel's volume change arises only from uptake or loss of fluid.

By insertion of Fick's law (equation \eqref{eq:flux_j}), the constitutive model for the concentration $\breve{c}$ (equation \eqref{eq:specific_mu}), and the isotropic form of the species mobility tensor $\mathbf{M}$ (equation \eqref{eq:mobility_isotropic}) into the balance of mass (equation \eqref{eq:fluid_balance_robin}), the latter can be expressed in terms of the chemical potential $\mu$ as primary variable:
\begin{equation}\label{eq:fluid_balance_mu}
   \frac{1}{\Lambda} \partial_t \mu + \frac{\chi}{\Lambda} \text{tr} \left( \partial_t \bm{\varepsilon} \right) + \frac{D}{k_B T} \text{div} \left( \text{grad} \left( \mu \right)\right) = 0, ~ \text{in} ~ \mathcal{B} \times I.
\end{equation}

The ﬁrst term on the right-hand side in equation \eqref{eq:fluid_balance_mu} represents the gel's dilation rate and plays an important role in the diffusion of the fluid in classic poroelastic materials, but can be neglected in our case due to the above considerations on hydrogels' incompressibility. The balance of mass \eqref{eq:fluid_balance_mu} can then be reduced further to
\begin{equation}\label{eq:fluid_balance_mu_final}
   \text{tr} \left( \partial_t \bm{\varepsilon} \right) = \frac{D\Omega}{k_B T} \text{div} \left( \text{grad} \left( \mu \right)\right). 
\end{equation}

Finally, the balance equations of linear momentum, equation \eqref{eq:linear_momentum}, and mass, equation \eqref{eq:fluid_balance_mu_final}, complemented by the constitutive equation for the stress, equation \eqref{eq:specific_sigma}, and the kinematic constraint, equation \eqref{eq:kinematic_cons}, constitute the \textit{specialized diffusion-deformation theory for small deformations superposed on a previously homogeneously swollen gel} \citep{Anand2020CMM_solids_Book}[Chapter 16.5], which reads as follows
\begin{problem}[Strong form]\label{eq:full_model}
    Given $\bar{\mathbf{u}},\bar{\mathbf{t}},\mu_{\infty}$ as boundary data and $\bm{\sigma}_0,\mu_0$ as initial data, find 
    $\bm{u}:\mathcal{B}\times I\to\mathbb{R}^d$ and $\mu:\mathcal{B}\times I\to\mathbb{R}$ such that
\begin{align}
    \text{div} (\bm{\sigma}) + \mathbf{b} = \bm{0}, & ~ \text{in} ~ \mathcal{B},\\
    \text{tr} \left( \partial_t \bm{\varepsilon} \right) = \frac{D\Omega}{k_B T} \text{div} \left( \text{grad} \left( \mu \right)\right), & ~ \text{in} ~ \mathcal{B} \times I,\\
    \mathbf{u} = \bar{\mathbf{u}}, & ~ \text{on} ~ \partial \mathcal{B}_{\mathbf{u}} \times I, \\
            \bm{\sigma} \cdot \mathbf{n} = \bar{\mathbf{t}}, & ~ \text{on} ~ \partial \mathcal{B}_{\bar{\mathbf{t}}} \times I, \\
            \bm{\sigma}\vert_{t = 0} = \bm{\sigma}_0, & ~ \text{in} ~ \mathcal{B} \times \lbrace T = 0 \rbrace , \\
          \text{grad} (\mu) \cdot \mathbf{n} = \alpha_R (\mu - \mu_{\infty}), & ~ \text{on} ~ \partial \mathcal{B}_{R}\times I, \\
        \mu\vert_{t = 0} = \mu_0, & ~ \text{in} ~ \mathcal{B}\times \lbrace T=0 \rbrace.
\end{align}
\end{problem}
For the numerical solution, we employ a normalized variant of Problem~\ref{eq:full_model}; see Section~\ref{sec:gels_theory}.

\section{2D square block - Full-order model convergence analysis}
\label{app:conv_analysis}

A computational convergence analysis is performed to investigate the robustness and computational cost of the monolithic approach. We focus on investigating the effect of mesh density and the time step size on the behavior of the implemented numerical algorithm. We aim to understand how these parameters influence the performance of the algorithm in solving the linear system of equations associated with the FEM discretization.

We measure the $L_2$ and $H_1$ errors in space for various mesh densities and time step sizes with respect to the highest fidelity solution. That is,
\begin{equation}
     \|\mathbf{u} - \mathbf{u}_{kh}\|_{L^2} = \sqrt{\int_{\Omega} (\mathbf{u}(T = 0.25)- \mathbf{u}_{kh}(T = 0.25))^2 \, dx},
\end{equation}
for the displacement, with $\mathbf{u}$ being a reference solution obtained using a very fine mesh and $\mathbf{u}_{kh}$ the obtained displacement obtained for either different mesh or time step sizes.

\begin{figure}[!htb]
\center{\includegraphics[width=0.9\textwidth]{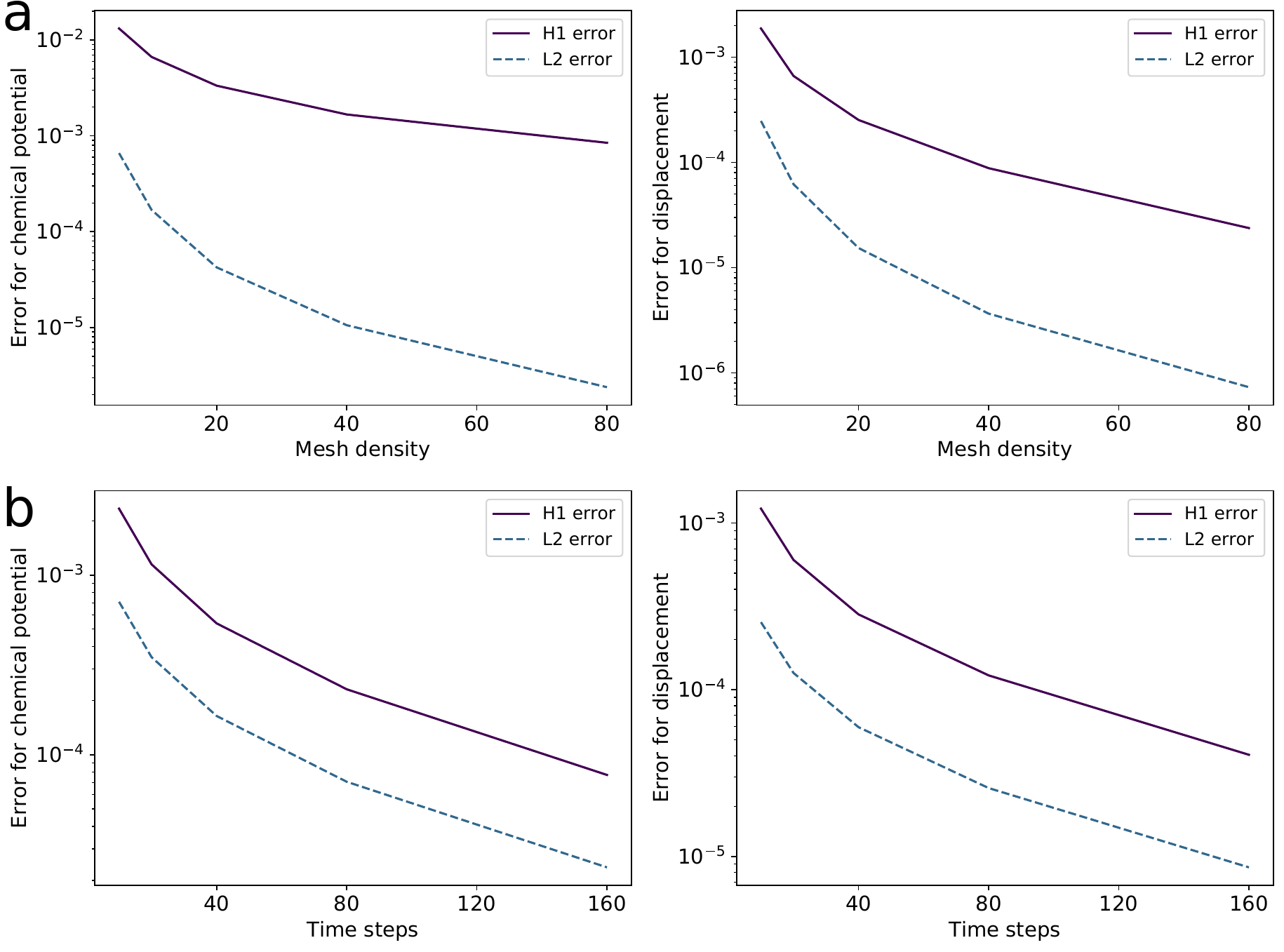}}
\caption{
FOM computational convergence analysis. For increasing \textbf{a.} mesh density (while fixing the time step size) and \textbf{b.} smaller time steps (while fixing the spatial mesh size). Thus, in the upper row, spatial convergence properties are measured, while in the bottom row, still $H^1$ and $L^2$ spatial evaluations are adopted, but with varying time step sizes, which measure the temporal convergence. As the first-order implicit Euler method is used, the same convergence orders for all four curves, namely chemical potential and displacements measured in both the $L^2$ norm and $H^1$ norm are observed and were expected.
}
\label{fig:FOM_conv}
\end{figure}

Figure~\ref{fig:FOM_conv} shows the $L^2$ and $H^1$ error norms, evaluated as both mesh size (Figure~\ref{fig:FOM_conv}\textbf{a}) and time steps (Figure~\ref{fig:FOM_conv}\textbf{b}) are refined. 
The errors are calculated against a benchmark of a fine mesh and small time steps. For the spatial dimension, the error analysis involves refining from a coarser $10 \times 10$ mesh to a finer $80 \times 80$ mesh. For the temporal dimension, the refinement process starts with $10$ time steps and increases to $160$, discretizing the dimensionless time interval more finely.
From Figure~\ref{fig:FOM_conv}, it is observed that in both cases the error keeps decreasing as mesh size and time steps refine.

We estimate the convergence order of the numerical method implemented with a well-known heuristic formula.
Let us denote the errors by $E_h$, $E_{h/2}$, and $ E_{h/4}$, where $h$ is the 
mesh size parameter as before. Under the assumption that our discretized problem has a convergence order of $p$, then each error should be roughly $(1/2)^p$ times the previous error. Therefore, we can estimate $p$ by taking the logarithm base $2$ of the error ratios as:

\begin{equation}\label{eq:conv_order}
    p = \dfrac{1}{\log (2)} \log \left( \left\vert \dfrac{E_h - E_{h/2}}{ E_{h/2} - E_{h/4}} \right\vert \right).
\end{equation}

The computational convergence analysis is detailed in Table~\ref{table:FOM_conv}. From finite element interpolation estimates for non-coupled problems and with optimal regularity of coefficients, we expect for polynomial shape functions of degree $s$ in the $L^2$ norm the order $s+1$ and 
in the $H^1$ norm $s$. In the temporal direction, the first order implicit Euler scheme has the order $r=1$, irrespectively of the spatial evaluation means. Thus, the temporal convergence analysis results are nearly optimal, as observed in Table~\ref{table:FOM_conv}. 
For the spatial analysis, for the non-coupled case with optimal regularity assumptions, we would expect for $s=1$ the order $2$ in the $L^2$ norm and order $1$ in the $H^1$ norm. 
Furthermore, we would expect for $s=2$ the order $3$ in the $L^2$ norm and order $2$ in the $H^1$ norm, and, finally, we would expect for $s=3$ the order $4$ in the $L^2$ norm and order $3$ in the $H^1$ norm. Specifically, we come close to these orders for the chemical potentials. For the displacements, we have slightly less optimal orders. However, our main assumption in these theoretical expectations was 
non-coupled partial differential equations with optimal regularity assumptions in the coefficients, the domain, the boundary, and the given problem data. Due to the nonlinear coupling, it is likely that the optimal orders are not achieved. Nonetheless, our findings certainly point in the right direction and clearly serve the purpose of the robust finite element verification of the full-order model.

\begin{table}[!ht]
\centering
\caption{FOM convergence analysis.}
\label{table:FOM_conv}
\begin{tabular}{lcccccc}
\toprule
 & & \multicolumn{4}{c}{\textbf{Convergence order}} \\ 
\cmidrule(lr){3-6}
 & & $\mathbf{u}$(degree=2) & $\mathbf{u}$(degree=3) & $\mu$(degree=1) & $\mu$(degree=2) \\ 
\midrule
\multirow{2}{*}{\textbf{Space}} & $L^2$ norm & 2.3222 & 3.4899 & 1.9623 & 2.9713 \\
 & $H^1$ norm & 1.6617 & 2.5822 & 0.9646 & 1.9801 \\
\midrule
\multirow{2}{*}{\textbf{Time}} & $L^2$ norm & 0.9462 & -- & 0.9631 & -- \\
 & $H^1$ norm & 0.9610 & -- & 0.9735 & -- \\
\bottomrule
\end{tabular}
\end{table}

\section{Reduced-order model convergence analysis}
\label{app:conv_analysis_ROM}

Tables~\ref{tab:POD_error_analysis} and \ref{tab:nestedPOD_error_analysis} refer to the error analysis performed with the testing set of the material parameters samples for the benchmark problem in Subsection~\ref{subsec:benchmark}.
The tables summarize the $L^1, L^2$, and $L^{\infty}$ error norms for an increasing number of basis functions.
We observe that the error decreases for both POD and nested-POD-based ROMs as the number of basis functions increases to six. After that, the marginal decrease in the error is negligible, showing that six basis functions are enough to retain the desired fraction of $\eta = 99.9999\%$ of the total energy in the FOM. Notably, the errors are very similar for both POD and nested-POD-based ROMs.
The reader should notice that these low errors were achieved by training the ROM with 30 parameter samples only. 
The good behavior of the errors reported in the tables demonstrates the good approximation obtained by the POD-based ROMs in the parameters space and the advantage of projection-based model-order reduction techniques, which encode the problem physics.

\begin{table}[!ht]
    \centering
    \caption{\textbf{Benchmark: POD-based ROM} -- norm-based error analysis for increasing number of basis functions.}
    \begin{tabular}{cccccccccc}
        \toprule
        r & \multicolumn{2}{c}{$L^2$ error ($\mathbf{u}$)} & \multicolumn{2}{c}{$L^1$ error ($\mathbf{u}$)} & \multicolumn{2}{c}{$L^{\infty}$ error ($\mathbf{u}$)} \\
        \cmidrule(lr){2-3} \cmidrule(lr){4-5} \cmidrule(lr){6-7}
        & mean & max & mean & max & mean & max \\
        \midrule
        1 & 0.006812 & 0.015478 & 0.005001 & 0.013313 & 0.021535 & 0.048306 \\
        2 & 0.001662 & 0.005001 & 0.001107 & 0.002850 & 0.007800 & 0.025559 \\
        3 & 0.000792 & 0.002991 & 0.000616 & 0.002698 & 0.003759 & 0.010280 \\
        4 & 0.000326 & 0.000843 & 0.000249 & 0.000763 & 0.001381 & 0.003892 \\
        5 & 0.000290 & 0.000434 & 0.000212 & 0.000364 & 0.001345 & 0.003955 \\
        6 & 0.000262 & 0.000358 & 0.000199 & 0.000341 & 0.000876 & 0.002091 \\
        7 & 0.000256 & 0.000289 & 0.000189 & 0.000272 & 0.000869 & 0.001986 \\
        8 & 0.000255 & 0.000280 & 0.000188 & 0.000254 & 0.000866 & 0.001827 \\
        \bottomrule
    \end{tabular}

    \vspace{0.5cm}

    \begin{tabular}{cccccccccc}
        \toprule
        r & \multicolumn{2}{c}{$L^2$ error ($\mu$)} & \multicolumn{2}{c}{$L^1$ error ($\mu$)} & \multicolumn{2}{c}{$L^\infty$ error ($\mu$)} \\
        \cmidrule(lr){2-3} \cmidrule(lr){4-5} \cmidrule(lr){6-7}
        & mean & max & mean & max & mean & max \\
        \midrule
        1 & 0.005766 & 0.014770 & 0.004850 & 0.012184 & 0.014942 & 0.054872 \\
        2 & 0.001787 & 0.004560 & 0.001450 & 0.003812 & 0.005110 & 0.022370 \\
        3 & 0.001043 & 0.003840 & 0.000821 & 0.003046 & 0.003746 & 0.018686 \\
        4 & 0.000479 & 0.001308 & 0.000363 & 0.001005 & 0.001709 & 0.005616 \\
        5 & 0.000415 & 0.001317 & 0.000331 & 0.000823 & 0.001307 & 0.007697 \\
        6 & 0.000395 & 0.001286 & 0.000305 & 0.000751 & 0.001326 & 0.007909 \\
        7 & 0.000386 & 0.001441 & 0.000302 & 0.000648 & 0.001180 & 0.011706 \\
        8 & 0.000385 & 0.001480 & 0.000299 & 0.000572 & 0.001102 & 0.013450 \\
        \bottomrule
    \end{tabular}
    \label{tab:POD_error_analysis}
\end{table}

\begin{table}[!ht]
    \centering
    \caption{\textbf{Benchmark: nested-POD-based ROM} -- norm-based error analysis for increasing number of basis functions.}
    \begin{tabular}{cccccccccc}
        \toprule
        r & \multicolumn{2}{c}{$L^2$ error ($\mathbf{u}$)} & \multicolumn{2}{c}{$L^1$ error ($\mathbf{u}$)} & \multicolumn{2}{c}{$L^{\infty}$ error ($\mathbf{u}$)} \\
        \cmidrule(lr){2-3} \cmidrule(lr){4-5} \cmidrule(lr){6-7}
        & mean & max & mean & max & mean & max \\
        \midrule
        1 & 0.006818 & 0.015437 & 0.005005 & 0.013272 & 0.021560 & 0.048486 \\
        2 & 0.001686 & 0.005047 & 0.001119 & 0.002872 & 0.007961 & 0.025928 \\
        3 & 0.000854 & 0.002979 & 0.000644 & 0.002658 & 0.004397 & 0.010913 \\
        4 & 0.000330 & 0.000828 & 0.000250 & 0.000750 & 0.001474 & 0.004087 \\
        5 & 0.000294 & 0.000457 & 0.000214 & 0.000345 & 0.001439 & 0.004179 \\
        6 & 0.000262 & 0.000346 & 0.000200 & 0.000333 & 0.000885 & 0.002060 \\
        7 & 0.000257 & 0.000283 & 0.000190 & 0.000261 & 0.000878 & 0.002051 \\
        8 & 0.000256 & 0.000280 & 0.000189 & 0.000250 & 0.000877 & 0.001933 \\
        \bottomrule
    \end{tabular}

    \vspace{0.5cm}

    \begin{tabular}{cccccccccc}
        \toprule
        r & \multicolumn{2}{c}{$L^2$ error ($\mu$)} & \multicolumn{2}{c}{$L^1$ error ($\mu$)} & \multicolumn{2}{c}{$L^\infty$ error ($\mu$)} \\
        \cmidrule(lr){2-3} \cmidrule(lr){4-5} \cmidrule(lr){6-7}
        & mean & max & mean & max & mean & max \\
        \midrule
        1 & 0.005777 & 0.014942 & 0.004859 & 0.012331 & 0.014927 & 0.055315 \\
        2 & 0.001802 & 0.004623 & 0.001463 & 0.003865 & 0.005119 & 0.022755 \\
        3 & 0.001069 & 0.003664 & 0.000850 & 0.002949 & 0.003742 & 0.017067 \\
        4 & 0.000482 & 0.001317 & 0.000366 & 0.001012 & 0.001705 & 0.005672 \\
        5 & 0.000418 & 0.001312 & 0.000335 & 0.000832 & 0.001303 & 0.007571 \\
        6 & 0.000395 & 0.001281 & 0.000305 & 0.000759 & 0.001320 & 0.007731 \\
        7 & 0.000386 & 0.001434 & 0.000302 & 0.000662 & 0.001181 & 0.011498 \\
        8 & 0.000385 & 0.001479 & 0.000300 & 0.000575 & 0.001107 & 0.013402 \\
        \bottomrule
    \end{tabular}
    \label{tab:nestedPOD_error_analysis}
\end{table}

\end{appendices}

\end{document}